\documentclass[amsmath, amssymb, english, aps, prb, twocolumn, 
reprint,floatfix, superscriptaddress,showpacs]{revtex4-1}

\usepackage[english]{babel}
\usepackage{graphics}
\usepackage{graphicx}
\usepackage{epsfig}
\usepackage{verbatim}
\usepackage[usenames,dvipsnames]{color}

\def\tg{t${}_{2g}$}
\begin{document}
\newcommand{\eg}{e${}_{g}$}
\newcommand{\note}[1]{{\bf[[NOTE: #1 ]]}}
\newcommand{\STO}[0]{Sr$_2$TcO$_4$}
\newcommand{\mach}[1]{\textcolor{red}{#1}}
\newcommand{\jm}[1]{\textcolor{Thistle}{#1}}
\newcommand{\lp}[1]{\textcolor{blue}{#1}}
\newcommand{\macomm}[1]{\textcolor{red}{[MACOMM: #1]}}

\title{Theoretical prediction of antiferromagnetism in layered perovskite Sr$_2$TcO$_4$}

\author{Alen~Horvat}
\affiliation{Jo\v{z}ef Stefan Institute, Jamova~39, Ljubljana, Slovenia}
\author{Leonid~Pourovskii}
\affiliation{Centre de Physique Th\'eorique, \'Ecole Polytechnique,
  CNRS, 91128 Palaiseau Cedex, France}
\author{Markus~Aichhorn}
\affiliation{Institute of Theoretical and Computational Physics, TU Graz, Petersgasse 16, Graz, Austria}
\author{Jernej~Mravlje}
\affiliation{Jo\v{z}ef Stefan Institute, Jamova~39, Ljubljana, Slovenia}

\begin{abstract}

We theoretically investigate magnetic properties of Sr$_2$TcO$_4$, a
4d transition-metal layered perovskite of the K$_2$NiF$_4$-type with
half-filled t$_{2g}$ states. The effect of local Coulomb repulsion
between the t$_{2g}$ orbitals is included within the
density-functional theory (DFT)+U and DFT+dynamical mean-field theory 
(DMFT) methods.  The DFT+DMFT predicts paramagnetic Sr$_2$TcO$_4$ to
be close to the Mott insulator-to-metal transition, similarly to the
cubic compound SrTcO$_3$.  The inter-site exchange interactions
computed within the DFT+DMFT framework point to a strong
antiferromagnetic coupling between the neighboring Tc sites within
the layer.  We then evaluate the N\'eel temperature $T_N$ within a
classical Monte Carlo approach including dipolar interactions, which
stabilize the magnetic order in the frustrated K$_2$NiF$_4$ lattice
structure. Our approach is applied to a set of layered and cubic
perovskites. The obtained $T_N$ are in fair agreement with
experiment. Within the same approach we predict $T_N$ of Sr$_2$TcO$_4$
to be in the 500-600\,K range.
\end{abstract}
\maketitle

\section{Introduction}

Recently, antiferromagnetism persisting to high temperatures has been
found in Tc$^{4+}$ perovskites SrTcO$_3$ and
CaTcO$_3$.\cite{rodriguez11,avdeev11} In 4d oxides the magnetism is
found less often than in more localized 3d oxides, hence strong
magnetic properties were not anticipated. As technetium is
radioactive, Tc perovskites have not been  intensely
experimentally investigated. Encountering a new class of compounds in
a familiar perovskite lattice with little  experimental information
provides a unique opportunity to test theoretical tools. In this paper
we theoretically investigate magnetic properties of layered perovskite
Sr$_2$TcO$_4$.

What makes Tc perovskites special among 4d compounds is
that they have a half-filled t$_{2g}$ shell. \cite{demedici11}
Several theoretical works~\cite{rodriguez11,franchini11, mravlje12}
discussed the role of more extended 4d orbitals, which give rise to a
large hybridization and small values of Hubbard interaction compared
to the 3d elements, and hence strong exchange interactions. Nevertheless, in addition to Refs.~\onlinecite{rodriguez11,franchini11, mravlje12} most of the recent work\cite{middey_sarma_12, wang12, dai14} recognized that electronic correlations remain sizable in Tc compounds, too; 
if the interactions were too small, the localized magnetic moments
would not establish. The key ingredient, that helps localization is
the Hund's rule coupling, \cite{georges_annrev13} that for the
half-filled shell increases the cost of charge excitations. \cite{demedici11}
 As a joint result of the Hund's rule coupling and
the more extended 4d orbitals, the Tc compounds are situated right at
the itinerant-to-localized transition \cite{demedici11} where the
N\'eel temperatures are maximal. \cite{mravlje12} Other 4d oxides do
not have half-filled shells \cite{georges_annrev13} and, consequently,
rarely exhibit ordered antiferromagnetism.

A layered Tc$^{4+}$ perovskite, Sr$_2$TcO$_4$, has been
synthesized,\cite{Goodenough70} too.  It crystallizes in the layered
body-centered tetragonal K$_2$NiF$_4$-type lattice, with the lattice
constants $a=3.902$\,\AA, $c=12.72$\,\AA, and space group
I4/mmm. \cite{Goodenough70,hartmann2012} To our knowledge, only its
basic crystal structure has been reported, but more detailed
measurements, in particular, the determination of the magnetic
structure, have not been performed to date.

From the similarity with the cubic SrTcO$_3$ one may expect sizable
magnetic interactions also in the layered
Sr$_2$TcO$_4$, \cite{middey_sarma_12} but the question of the long-range 
ordering in layered systems is more subtle,\cite{mermin1966} in particular for the lattices of the  K$_2$NiF$_4$-type. Usually, in quasi-2D the transition temperatures are not suppressed so much with respect to the 3D lattices. The magnetic susceptibility in 2D increases exponentially on cooling down.\cite{chakravarty89} The usual argument then implements  the inter-plane exchange interaction $J_\perp$ at the mean-field
level,\cite{scalapino75,schulz96} which leads to only
logarithmic suppression of the ordering temperature 
$T_c\sim  T_c^{3\mathrm{D}} / \log( b J/ J_{\perp})$ 
with diminishing $J_{\perp}$ ($b$~is model dependent number).\cite{chakravarty89, yasuda05}  In
the present case of the body-centered K$_2$NiF$_4$ structure, however, the
frustration suppresses the effective coupling between layers
completely, provided the in-plane order is checkerboard
antiferromagnetic. When this is the case (as it turns to be in Sr$_2$TcO$_4$), each spin is equally coupled
to the same number of oppositely oriented spins in the layer above
(and likewise below), hence the net inter-plane coupling cancels out.
Nevertheless, the antiferromagnetic order is experimentally found also
in lattices of this kind, for instance in K$_2$NiF$_4$\cite{lines67} and
Rb$_2$MnF$_4$.\cite{zhou07} The reason can be traced to the magnetic anisotropies
that originate in the dipole-dipole
interactions.\cite{lines67, cowley_birgenau_1977, christianson_birgenau_2001, zhou07}
If a pair of
magnetic moments is forced to point in opposite directions by a strong
antiferromagnetic exchange, the dipole-dipole energy is reduced when
the moments are oriented perpendicular to the line connecting them. In
a planar configuration with antiferromagnetic arrangement of nearest
neighbors, the moments will prefer to point in a direction perpendicular to the
plane. Such effective anisotropy in the presence of a long in-plane
correlation length is sufficient to stabilize the order already in a
single layer. Once the order is established in 2D, the long-range
order in 3D follows due to any non-vanishing next-nearest layer coupling.

In the present work we study theoretically the electronic structure
and magnetic properties of Sr$_2$TcO$_4$. 
 We calculated the exchange
interactions using a linear-response approach based on DFT+DMFT and
found a large antiferromagnetic coupling between the nearest
neighboring Tc moments.  In order to determine the N\'eel temperature, we employed
a classical spin Monte Carlo technique with dipole-dipole interactions
included.\cite{landau_binder, Roesler1999} We tested our approach on a
set of layered and cubic perovskites, for which the experimental
values of exchange parameters are known.  This allows us to
estimate systematic errors of our theoretical exchange
interactions. With a correction for this systematic error included,
we predict the transition temperature of Sr$_2$TcO$_4$ to be about $500$\,K.  

The paper is structured as follows. In Sec.~\ref{sec:methods} we
describe our theoretical approach. In Sec.~\ref{sec:results} we
present our DFT+U and DFT+DMFT results from which we infer the magnetic
ordering. In Sec.~\ref{sec:mc} we report the results of the calculated
exchange interactions and transition temperatures for a set of
perovskites including the predicted value for Sr$_2$TcO$_4$. The
relevance of effects that were not included in our results is
discussed in Sec.~\ref{sec:disc}. Appendix~\ref{app:leonid} gives
further details of the calculation of exchange interactions.

\section{Methods}

\label{sec:methods}

\subsection{DFT+U}
We have employed a rotationally invariant DFT+U implementation of the
Wien2k package,\cite{Wien2k} and used $U_{\rm eff}=U-J=2.04$\,eV as the
value for the Coulomb repulsion.
\footnote{DFT+U calculations were carried out using the implementation of the Wien2k package within GGA approximation. 
The correlated orbitals used in this implementation are different from those employed in the TRIQS package, hence, we employed 
slightly different values of U in our DFT+DMFT and DFT+U calculations of Sr$_2$TcO$_4$ }
The double-counting correction
term\cite{Anisimov1993} was taken in the fully-localized
limit, and we used the generalized-gradient approximation (GGA)\cite{PBE96} as approximation to the exchange
correlation potential.  The Brillouin zone (BZ) integration was
carried out with 1000 $\mathbf k$-points in the full BZ, which
corresponded to 56 (150) $\mathbf k$-points in the tetragonal
(orthorhombic) irreducible BZ.  The muffin-tin (MT) radii were fixed at
1.81, 2.1 and 1.6\,a.u. for Tc, Sr and O, respectively, in all total
 energy and structural relaxation calculations.

\subsection{DFT+DMFT}
The influence of electronic correlations was also investigated within
the DFT+DMFT approach. We use the efficient implementation of this
method as provided by the TRIQS
package.\cite{triqs,aichhorn09,aichhorn11} Based on DFT calculations
within the local-density approximation (LDA) using Wien2k, we
construct Wannier functions for the t$_{2g}$ orbitals, which serve as a
basis for the DMFT calculations. The solution of the DMFT quantum
impurity problem was done by a continuous-time quantum Monte Carlo
(CTQMC) method in hybridization
expansion\cite{gull11,werner2,boehnke11} including full rotational invariant interactions.\cite{parragh12}  
We use the same interaction
values $U=2.3$\,eV and $J_H=0.3$\,eV as previously for
SrTcO$_3$.\cite{mravlje12} We performed paramagnetic and magnetic
calculations. For the magnetic calculation, we used unit cell that
could accommodate the G-type ordering pattern, which is a checkerboard
AFM in the plane and FM stacking in $c$ direction in the unit cell
containing two oppositely oriented Tc$^{4+}$ spins. In the
paramagnetic calculation, standard tetragonal unit-cell was used and
the self-energies were spin-symmetrized after each iteration of the DMFT loop.


\subsection{Calculation of exchange interactions}
\label{sec:magint}

In order to calculate the magnetic transition temperature we define an effective Hamiltonian associated with magnetic degrees of freedom. 
In Mott insulators this Hamiltonian is known to be well approximated by the quantum Heisenberg form
\begin{equation}
  \label{eq:spin_hamiltonian0}
  \mathcal{H} = -\frac{1}{2}\sum_{i\neq j}J_{ij}\mathbf{\hat{S}}_i\cdot\mathbf{\hat{S}}_j,
\end{equation}
where $\mathbf{\hat{S}}$ are  spin operators and $J_{ij}$ are inter-site 
exchange interactions. We denote $J_{i,j}=J$ for nearest neighbor interactions.

The inter-site exchange interactions were extracted from the results
of DFT+DMFT calculations within the Hubbard-I\cite{hubbard_1}
approximation (DFT+HubI) using an approach similar to the DFT-based
linear response techniques introduced in
Refs.~\onlinecite{Liechtenstein1987,Bruno2003,Ruban2004} (and generalized to
DFT+DMFT in Refs.~\onlinecite{Katsnelson2000,Wan2006}).  Similarly to those
techniques, our approach is based on evaluating a linear response to
simultaneous magnetic fluctuations on two neighboring sites. The
standard ``Lichtenstein interactions"
\cite{Liechtenstein1987,Bruno2003} and their DFT+DMFT generalizations
are derived by considering simultaneous tilting of two moments in a
magnetically-ordered ground state. In contrast, we compute inter-site
exchange interactions from a local-moment paramagnetic state
(described within Hubbard-I), which is advantageous when the
ground-state magnetic order is not known. More details follow in Appendix~\ref{app:leonid}.

%

\subsection{Mapping to the classical spin Heisenberg model}

We approximate the quantum Heisenberg model Eq.~\eqref{eq:spin_hamiltonian0} with the classical one:
\begin{equation}
  \label{eq:spin_hamiltonian}
  \mathcal{H}_{\mathrm{c}} = -\frac{1}{2}\sum_{i\neq j}J_{ij}\alpha\mathbf{S}_i\cdot\mathbf{S}_j.
\end{equation}
$\mathbf{S}_i$ are three-dimensional vectors  of length unity and constant $\alpha$ 
keeps track of the length of the quantum spin. It has to be
adjusted in a way that the classical model gives the best
approximation to the quantum model. 

As $\hat{S}^2=S(S+1)$ (with $S=3/2$ in the ground state of the Tc ion) suggests $\alpha_1=S(S+1)$ should be used.
Indeed, largest deviations from the classical results are expected for
$S=1/2$ and even there the choice $\alpha_1$ gives for the
simple cubic lattice in 3D results that differ from the quantum ones
only by about $15\%$.\cite{Rushbrooke1963, yasuda05, oitmaa04}

One can rationalize also a different choice of $\alpha$. Consider a pair
of magnetic moments and compare the energy of parallel and
anti-parallel configurations. For quantum spins, the energy difference
is $\Delta E = J S(2S+1)$, which follows from comparing the
expectation value
\begin{equation}
  J\langle \hat{\mathbf{S}}_i\cdot \hat{\mathbf{S}}_j\rangle =  J [S'(S'+1)-S_i(S_i+1)-S_j(S_j+1)]/2,
\end{equation}
in the singlet
configuration (where the spin of the pair is $S'=0$) with the 
triplet configuration ($S'=2S$) of the pair. For the classical model, the corresponding energy difference is simply $2J_\mathrm{c} $ (denoting the
classical exchange as $J_\mathrm{c}$). Comparing the two results, leads to a
choice $\alpha_2 = S(S+1/2)$. 
A third choice $\alpha_3 = S^2$ is also
used in the literature. \cite{Martin1997, archer11}  

The experiment suggests that the choice $\alpha_1$ is the best
one. Namely, in several cases, where the exchange constants have been
measured experimentally, the calculated values of transition
temperatures agree best with the observed ones if $\alpha_1$ is used,
see Sec.~\ref{sec:results} and Ref.~\onlinecite{zhou07,archer11}. (Note, that in Ref.~\onlinecite{archer11}, experimental exchange constants have been rescaled according to $\alpha_3$.)

However, quite universally the numerical approaches overestimate the
exchange interactions.\cite{Wan2006,franchini11} We find this holds
for our results, too. In such cases, a smaller choice of $\alpha$ than
$\alpha_1$ improves the agreement with the experiment.  Hence, in the
following paper we give our results for all of the three mappings,
together with the known experimental results for easy comparison.

To correctly describe magnetic properties of the layered perovskites
one needs to take into account additionally long-range dipolar
interactions and/or single-ion anisotropy as will discussed in more
detail later.

\subsection{Monte Carlo simulations \label{mcsim}}

We describe the cubic perovskites with the isotropic Heisenberg
model Eq.~\eqref{eq:spin_hamiltonian}. We implemented two algorithms: the
modified Wolff cluster algorithm,\cite{Roesler1999} which is
applicable whenever spin frustration is weak, and the
Metropolis-Hasting algorithm with over-relaxation.\cite{landau_binder}
For the data described in the paper, the two
algorithms gave consistent results.


For layered perovskites of the K$_2$NiF$_4$-type, the dipole-dipole
magnetic interactions are crucial to explain finite N\'eel
temperatures.  Therefore, we add to the short-range Heisenberg-type
classical Hamiltonian Eq.~\eqref{eq:spin_hamiltonian} a contribution from
long-range dipolar interactions. The resulting total Hamiltonian reads
\begin{equation}
  \label{eq:spin_hamiltonian1}
  \mathcal{H}' = \mathcal{H}_{\mathrm{c}}+  \frac{\tilde{\mu}S^2}{2}\sum_{i\neq j} r_{ij}^{-3}\left[\mathbf{S}_i\cdot\mathbf{S}_j - 3(\mathbf{S}_i\cdot\hat{\mathbf{r}}_{ij}) (\mathbf{S}_j\cdot\hat{\mathbf{r}}_{ij})\right].
\end{equation}
We denote $\tilde{\mu} = (g\mu_B)^2\mu_0/(4\pi) = 0.214$\,meV\AA$^3$,
$g\approx 2$  is the Land\'e factor, $\mu_B$ is the Bohr magneton and
$\mu_0$ is the permeability constant. The dipolar term scales as
$S^2$, since the energy of a spin with a magnetic moment $g\mu_B S$
in a dipolar field generated by a spin with the same magnetic moment
is proportional to $S^2$.\cite{zhou07} 

Magnetic ions in the K$_2$NiF$_4$ structure form a body-centered
tetragonal unit cell.  Energy contributions of spins in the center of
the unit cell to a spin in the corner of the unit cell cancel due to
frustration.  Hence we disregard the central ions and the remaining
ions form a simple cubic tetragonal unit cell.  Two layers of spins
with long-range dipole-dipole interactions
Eq.~\eqref{eq:spin_hamiltonian1} and with open boundary conditions in
$z$ direction are simulated as in Ref.~\onlinecite{zhou07}.
Dipole-dipole interactions were included up to fifth nearest
neighbor. We checked that the results do not change if 
the range of the dipole-dipole interaction is increased further.



\section{Results}
\label{sec:results}
\subsection{DFT+U}
\label{dft_u}
We have considered three kinds of magnetic ordering: 
ferromagnetic (FM) ordering as well as A-type and G-type antiferromagnetic
(AFM) ordering. The A-type AFM structure has the tetragonal P4/mmm space group
with the Tc moments aligned ferromagnetically within (001) planes and
antiferromagnetically between the neighboring planes.  The G-type
structure has a doubled unit cell of orthorhombic Cmmm space group,
with the moments of Tc in the center of the $ab$ face being
anti-parallel to those of Tc in the vertices.

The calculated total energy of the AFM G-type phase at the
experimental lattice parameters\cite{Goodenough70} is 0.41\,eV below
those of the FM and 
AFM A-type ones. The total energy of the A-type phase is just 5\,meV
below the FM's one.  Because the moments of nearest neighbor Tc ions are
aligned ferromagnetically in the FM and AFM A-type structures, but
antiferromagnetically in the G-type AFM one, we conclude that a
very strong nearest-neighbor AFM coupling is by far the most
significant magnetic interaction in this system.  Longer-range
 interactions are weaker. For example, the second
nearest-neighbor interaction (the one between Tc moment in the corner and
 in the center of the  tetragonal cell) 
 gives opposite sign
contributions to the energies of the FM and A-type phases that are
almost degenerate in the present case.

\begin{figure}
  \centering \includegraphics[width=0.90\columnwidth]{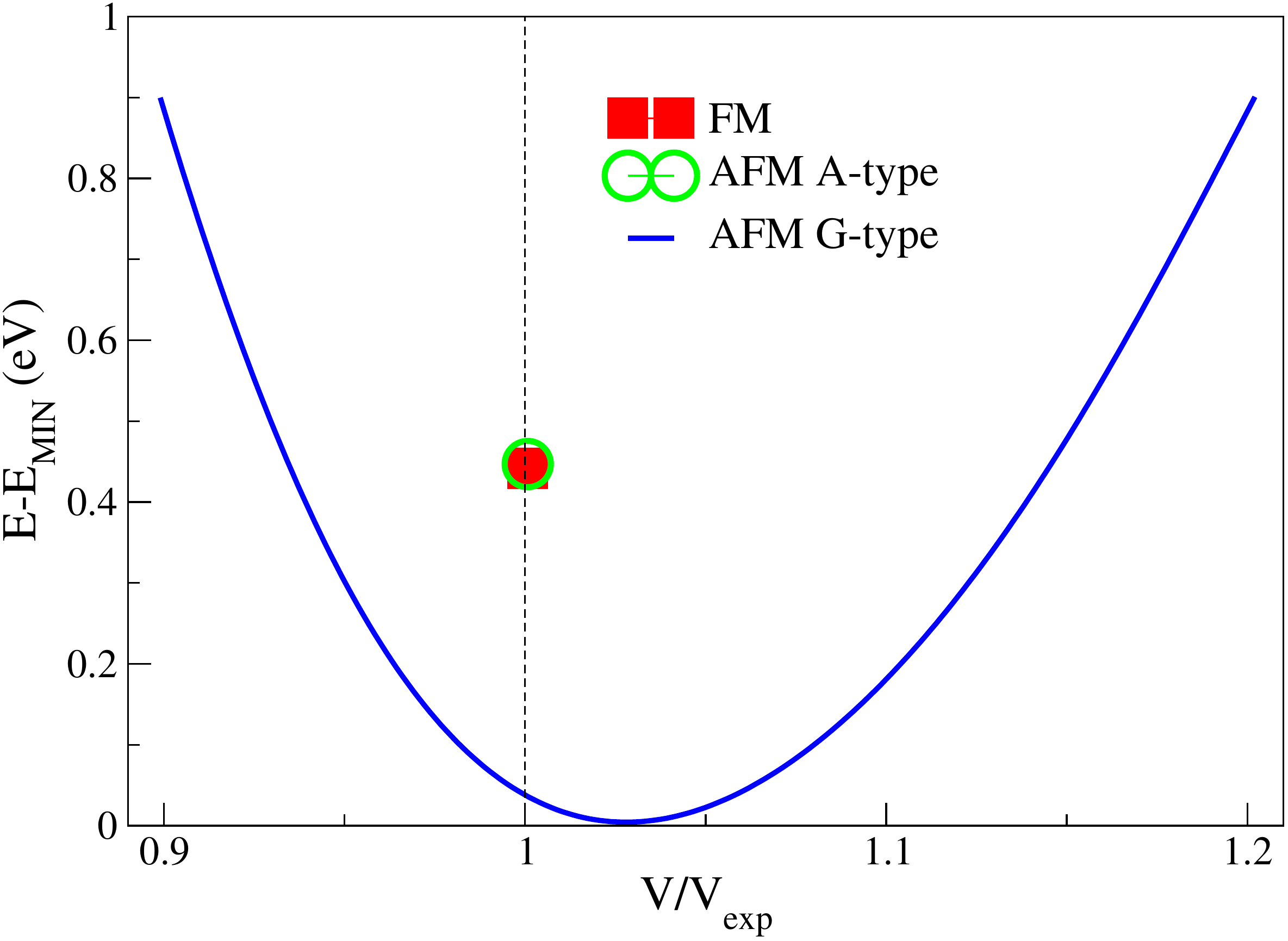}
  \caption{\label{fig:enr_LDA_U}%
    Solid line: the total energy vs. volume for the fully-relaxed
    AFM G-type structure of Sr$_2$TcO$_4$. The square and the circle
    indicate the total energies of the FM and AFM A-type structures, respectively. } 
\end{figure}

The G-type AFM is clearly the lowest-energy structure among all
 considered, and its large stabilization energy with respect to
the other two phases can hardly be offset by lattice relaxations.
Therefore, we performed lattice structure optimization for the AFM
G-type phase only.  We have found that orthorhombic distortions due to
the G-type ordering are negligible, and performed the full optimization
of the $a(=b)$, $c$ and internal coordinates.  As one may see in
Fig.~\ref{fig:enr_LDA_U} the theoretical equilibrium volume is 3\%
larger than the experimental one due to the usual tendency of GGA
towards a volume overestimation. The DFT+U equilibrium lattice
parameters $a(=b)$ and $c$ of the Cmmm structure are 5.61\,\AA \ and
12.68\,\AA, respectively. There are two distinctive $z_{\rm Sr}$ and three
distinctive $z_{\rm O}$ in the Cmmm structure due to symmetry lifting.
However, the difference between the optimized values of the
corresponding $z$ parameters is very small (below 0.1\%) and the
resulting values are $z_{\rm O_1(O_2)}=0.162$, $z_{\rm O_3}=0.50$, and
$z_{\rm Sr_1(Sr_2)}=0.353$. The latter two values are in almost
perfect agreement with experimental data,\cite{hartmann2012} only
$z_{\rm O_1}$ is slightly smaller in our calculation. 
Very similar values are obtained already at the 
non-magnetic LDA level. 

\begin{figure}[t]
  \centering
  \includegraphics[width=0.95\columnwidth]{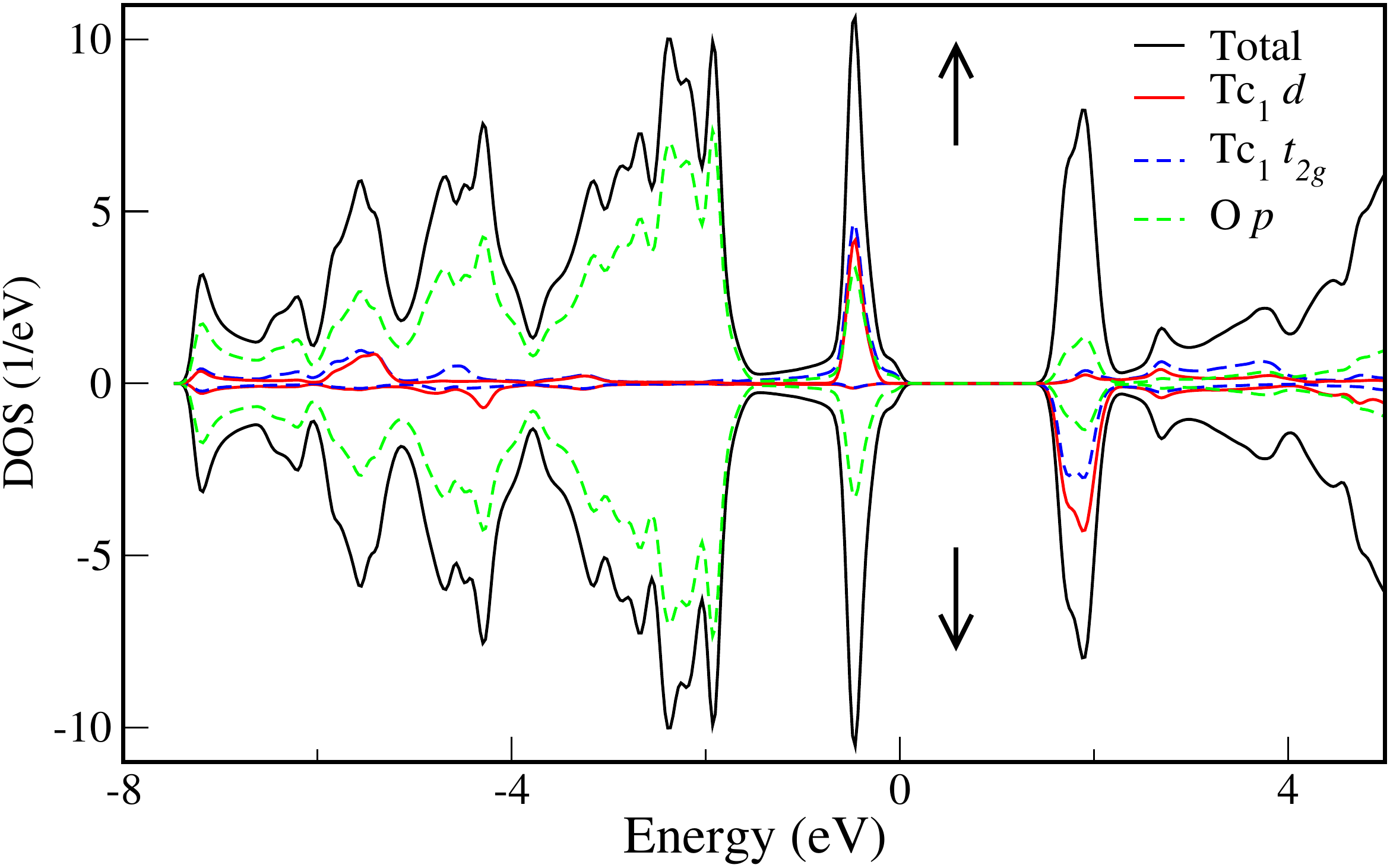}
  \caption{\label{fig:LDAU_DOS}%
Total and partial Tc$_1$ d,
    t$_{2g}$ and O p DOS for the G-type AFM structure obtained by
    the GGA+U method for the theoretical equilibrium lattice
    structure. 
}
    
\end{figure}

The density of
states (DOS) of the G-type structure calculated within DFT+U
(Fig.~\ref{fig:LDAU_DOS}) shows that the Tc t$_{2g}$ band is close
to full polarization, with almost no occupied t$_{2g}$ states
in the minority channel. As the t$_{2g}$ states extend also to the
oxygen orbitals where the contribution to the spin density in the
antiferromagnetic configuration cancel, the moment relevant to neutron
measurements can be expected to be smaller. One may try to estimate
this moment by looking at the spin polarization within MT spheres,
which gives 1.82\,$\mu_B$. As the compound is quite itinerant, some of
the spin density extends outside of the MT sphere and this value has
to be considered as the lower bound on the magnetic moment. 

One may also see that DFT+U gives insulating behavior for Sr$_2$TcO$_4$, with the gap of
about 1.4\,eV.

\subsection{ DMFT results}
For the relaxed structure of Sr$_2$TcO$_4$ we performed DFT+DMFT calculations.  In a
paramagnetic state we find that the metal-to-insulator transition
happens close to $U=2.3$\,eV, $J=0.3$\,eV, with these parameters
corresponding already to an insulating solution, while SrTcO$_3$ is
for the same parameters a paramagnetic metal. The difference between
the two materials is due to a smaller bandwidth of the
quasi-two-dimensional compound Sr$_2$TcO$_4$.

\begin{figure}
  \centering
  \includegraphics[width=0.95\columnwidth]{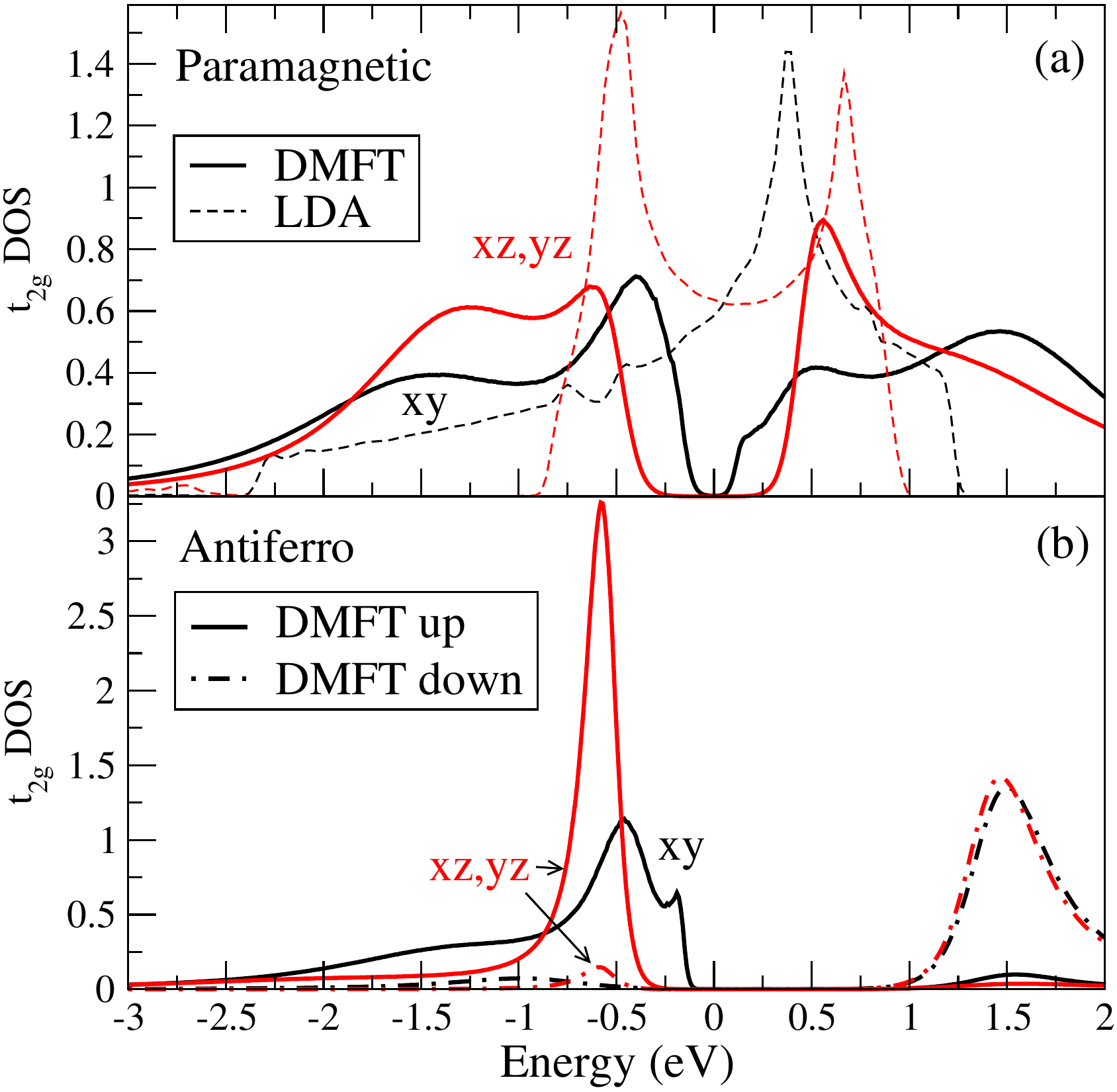}
  \caption{\label{fig:DOS}%
    Orbitally resolved DOS for \tg-states in Sr$_2$TcO$_4$
    calculated from LDA+DMFT at temperature $T=290$\,K, and from LDA. (a) Paramagnetic
    calculation. (b) Antiferromagnetic calculation.
  }
\end{figure}

On Fig.~\ref{fig:DOS} we plot the orbitally projected DOS for
Sr$_2$TcO$_4$ as calculated within DFT+DMFT using Maximum Entropy for
analytical continuation.\cite{Beach2004}  One can observe that
especially for the $xy$ orbital, the gap is very small, demonstrating
that for the selected parameters, the compound is very close to the
insulator-to-metal transition. On Fig.~\ref{fig:DOS} (a), besides the
paramagnetic DMFT DOS additionally the orbitally projected LDA DOS
in the paramagnetic state are plotted. These show that bands spanned
by the $xz,yz$ orbitals are narrower, which explains the occurrence of
a broader gap in the $xz,yz$ DMFT DOS. 

 In passing we note that stronger correlations in $xz,yz$ orbitals is
 opposite to what one finds in isostructural 4d$^4$ compound
 Sr$_2$RuO$_4$, where the strongest mass renormalization has been
 found in the $xy$ orbital, due to proximity to the van-Hove
 singularity.\cite{mravlje11} Sr$_2$TcO$_4$, being half-filled, is
 dominated by the proximity to the Mott transition, and the van-Hove
 singularity is further from the Fermi level, which both contribute to
 the fact that standard argument which associates the more narrow band
 with stronger correlations applies.

The fact that the compound is close to the metal-to-insulator
transition in the paramagnetic state points to strong tendencies,
as was discussed for 
SrTcO$_3$.\cite{mravlje12} Actually, if in DMFT one allows for
the magnetic ordering one gets a very similar behavior as in the case
of SrTcO$_3$ and similar DMFT ordering temperature about 2000\,K.
On Fig.~\ref{fig:DOS}(b) we plot the spin and orbitally resolved DMFT
DOS calculated well in the antiferromagnetic state at $T=290$\,K. Comparing to panel~(a), one can
notice that a bigger gap is opened by the onset of magnetic
order. The size of the gap is comparable to the one found DFT+U
(Fig.~\ref{fig:LDAU_DOS}). Conversely, in pure local spin density approximation (LSDA)  the gap in the
antiferromagnetic state is very small and only about 0.05\,eV.

The large ordering temperatures found in DMFT calculations should be
taken as pointing to strong magnetic correlations but cannot be
trusted quantitatively.  Furthermore, in quasi-2D these fluctuations become
more important than in 3D, hence even relative comparison to the
values found in SrTcO$_3$ is meaningless.

One can, however, assume that the ordering takes place and calculate
from DMFT the value of the ordered magnetic moment at low
temperatures. Employing the same scheme as for
SrTcO$_3$,\cite{mravlje12} i.e., calculating the moment in the set of
localized d-p Wannier functions, results in a moment of 2.1\,$\mu_B$. 
Again, this is significantly smaller than the
saturation value of 3\,$\mu_B$. The reason is partly due to covalence
and partly due to charge fluctuations, which both arise from strong
hybridization of Tc d with oxygen p states.

To investigate the ordering temperature itself one needs to employ
other approaches that are more suitable for quasi-2D systems.  We
discuss this  next.

\subsection{Exchange interactions and results of Monte Carlo simulations}
\label{sec:mc}

In order to obtain theoretical exchange interactions we have employed two approaches: First, the one  
described in Sec.~\ref{sec:magint}, and second, we also extracted nearest-neighbor $J_\mathrm{DFT+U}$ from the difference of
GGA+U total energies between ferromagnetic and AFM structures (we chose AFM cubic and tetragonal structures with all nearest-neighbor transition metal sites having opposite
spin directions).

In Table~\ref{tab:sc} we list our calculated and known experimental
values of the spin exchange interactions, which were obtained from
neutron and Raman scattering experiments as described in
Refs.~\onlinecite{jongh74, fleury68}. 
To our knowledge, only the exchange interactions for the first nearest
neighbors have been measured experimentally in the compounds under consideration,  
except for K$_2$NiF$_4$, where the exchange coupling for the next-nearest neighbors is $J'_\mathrm{exp}= 0.5$\,meV. 
One may notice that Sr$_2$TcO$_4$  has exceptionally  strong exchange couplings in
comparison with other layered perovskites.

\begin{table}
\centering
\caption{\label{tab:sc} Calculated exchange interactions compared to the experimental values from the literature.}
    \begin{tabular}{c| c c| c | c }
compound          &$J$\,[meV]  &$J'$\,[meV]    &  $J_\mathrm{DFT+U}$\,[meV] & $J_{\mathrm{exp}}$\,[meV]   \\ \hline
KNiF$_3$       &$-12.7$      &$-0.13$        &     $-7.7$               & $-8 $~\cite{jongh74}            \\
K$_2$NiF$_4$   &$-13.5$      &$-0.13$        &     $-8.4$               & $-8.6 $~\cite{jongh74}         \\
Rb$_2$MnF$_4$  & -           & -             &     $-1.1$               & $-0.65 $~\cite{jongh74}        \\
SrMnO$_3$      &$-8.1 $      &$-0.76 $       &     $-9.1$               & $-4.14 $~\cite{bouloux81}        \\
Sr$_2$MnO$_4$  & $-10.27$    & $-0.83$       &     $-12.4$               & $-6.89 $~\cite{bouloux81}        \\
SrTcO$_3$      &$-28.2 $     &$ -0.9 $       &     $-32.8$               &     -            \\
Sr$_2$TcO$_4$  &$-35.6 $     &$ -0.9 $       &     $-45.1$               &     -             \\
   \end{tabular}
\end{table}

Comparing the theoretical values to the measured ones, one notices that
the two approaches behave differently.  DFT+U works well for the
localized compounds; for the strongly localized Ni fluorides the
calculated exchange interactions are in close agreement to
experiment. The DFT+U estimates become progressively worse when the
localization becomes weaker and the Mott insulator-to-metal
transition is approached. 

The DFT+HubI approach, on the other hand, gives values that are
approximately 50\% larger than the experimental values irrespectively
of the vicinity of the Mott transition point.\footnote{A significant
  overestimation of the inter-site exchange within an approach based
  on Hubbard-I has been previously pointed out in
  Ref.~\onlinecite{Wan2006}.  Apparently, neglecting hybridization
  effects in calculation of the impurity Green's function is a rather
  crude approximation even for strongly-localized Mott insulators.}
For ``ionic'' Ni fluorides, less localized Mn perovskites, and most
itinerant SrTcO$_3$ one observes similar relative overestimate. As our
aim is to predict the transition temperature and furthermore for the
compound that is close to the Mott transition, we  employed the DFT+HubI values for $J$ Monte Carlo calculations described in the following.

The critical temperatures computed within the classical Monte Carlo approach for the set of  cubic and layered
perovskites are reported in Table~\ref{tab00}. In the first four columns we present
results of numerical simulations and in the last column we list the
known experimental values of  critical temperatures. In the first column 
we list calculated critical temperatures (denoted by $T_0$) for experimentally determined exchange
interactions 
 and mapping $\alpha_1=S(S+1)$. In our
evaluation of the critical temperature for experimental interactions, we calculated $J'$ (when it was not available from
experiment) by multiplying $J_\mathrm{exp}$ with the ratio $J'/J$ of theoretical
values. For layered perovskites, we added the long-range dipolar interaction as described in Sec.~(\ref{mcsim}).
Results in columns $T_0, T_\mathrm{exp}$ are in very good agreement, which supports the choice of $\alpha_{1}$ in the quantum to classical mapping.

\begin{table}
\caption{\label{tab00} In column $T_0$ we list calculated critical temperatures for experimental values of exchange interactions $J$. $T^{(\alpha_1)}$, $T^{(\alpha_2)}$, $T^{(\alpha_3)}$ are critical temperatures for theoretically calculated exchange interactions $J$, and $T_{\mathrm{exp}}$ are experimentally measured critical temperatures.}
    \begin{tabular}{c|c|c|c|c|c}
    compound & $T_0$\,[K]         & $T^{(\alpha_1)}$\,[K]        & $T^{(\alpha_2)}$\,[K]&  $T^{(\alpha_3)}$\,[K] & $T_{\mathrm{exp}}$\,[K]\\\hline
    KNiF$_3$      & 259              & 410      &   307      & 205         & 253~\cite{lines67}  \\
    K$_2$NiF$_4$  & 105              & 166      &   124      & 83          & 97.1~\cite{birgeneau70} \\
    Rb$_2$MnF$_4$ & 39.8             & -        &    -       & -            & 38.5~\cite{zhou07} \\
    SrMnO$_3$     & 259 & 370      &  296        & 222         & 260 ~\cite{takeda74}  \\
    Sr$_2$MnO$_4$  & 162             & 212      &  170        & 127         & 170 ~\cite{bouloux81}  \\
    SrTcO$_3$      & -            & 1610            & 1286     & 965         & 1023~\cite{rodriguez11} \\
    Sr$_2$TcO$_4$  & -            & 720             &  630     & 430         & -      \\
    \end{tabular}
\end{table}

In columns $T^{(\alpha_1)}$, $T^{(\alpha_2)}$, $T^{(\alpha_3)}$ we present critical temperatures for
theoretical values of exchange interactions and mappings $\alpha_1=S(S+1),
\alpha_2=S(S+1/2), \alpha_3=S^2$, respectively.  

From Table~\ref{tab00} it is obvious that the calculations of the
exchange interactions are not precise enough to allow for direct
prediction of the transition temperature.  However, the deviations are
systematic and therefore we can estimate from joint theoretical and
experimental data the transition temperature of
Sr$_2$TcO$_4$. 
We  extrapolate from the comparison between the theory and experiment
on  similar compounds (that furthermore all have $S=3/2$), in order
that systematic error (the ``correct'' choice of $\alpha$) drops out.
It is convenient to introduce the notation
\begin{equation}
T_x(a;b) = T^{(\alpha)} (a) \frac{T_\mathrm{exp}(b) } {T^{(\alpha)} (b) }
\end{equation}
that describes the predicted temperature $T_x$ for compound $a$ corrected
by the ratio of experimental and theoretical values evaluated on a
compound $b$.  Using that notation, one has $T_x$(Sr$_2$TcO$_4$;
SrTcO$_3$) = 456\,K and $T_x$(Sr$_2$TcO$_4$; Sr$_2$Mn$_4$) = 576\,K.
Averaging those two results gives 500\,K with 50\,K errorbar. Using
other entries of Table~\ref{tab00} for the reference compound $b$ does
not change the result significantly.  We also note that for all the
investigated compounds, the measured transition temperatures
$T_\mathrm{exp}$ are between $T^{(\alpha_2)}$ and $T^{(\alpha_3)}$,
therefore these two latter values can be used as the (somewhat
rougher) upper and lower limit for the predicted transition
temperature within our approach, too.


\section{Discussion}
\label{sec:disc}


Another reason for the magnetic ordering in the layered perovskites of
the K$_2$NiF$_4$ type is the single-ion
anisotropy\cite{lines67,birgeneau70} that originates in the
combination of spin-orbit interactions and the tetragonal crystal field. 
Whereas spin-orbit interaction is sizable in 4d
oxides its effects are expected to be less important here because 
in the present case of half-filled t$_{2g}$ orbitals the total orbital
momentum vanishes.
 Nevertheless, we calculated the ordering temperature also in the presence of 
 additional single-ion magnetic anisotropy and found that its effects are not  important (a $100\%$ increase of
 anisotropy above the value that is already effectively present due
 to the dipole-dipole interactions causes only $3\%$ increase of the critical
 temperature).

We investigated the  effects of the direct next-nearest layer coupling, too. Within our approach we estimate that this coupling is
very small, i.e. $\approx0.01$\,meV. Including this interaction
increases the transition temperature by about $50$\,K, from which follows our final prediction 550\,K with 50\,K errorbar. 

Finally, the orthorhombic distortion (that has not been reported
experimentally so far, and hence we did not include it), might lead
also to significant reduction of the frustration of the magnetic
couplings between the layers. This effect would make the magnetic
transition more of the 
La$_2$CuO$_4$ kind,\cite{kastner_rmp_1998} and would increase the transition temperature
further. Our DFT+U simulations, Sec.~\ref{dft_u}, predict that sizable orthorhombic distortions do not occur.  If such distortions were realized in the real structure nevertheless, then our results
should be taken as an estimate of the lower bound of the
transition temperature.

\section{Conclusions}
\label{sec:conc}

In summary, using a newly developed combination of theoretical
methods that we tested on a set of cubic and layered perovskites, we
investigated properties of the single-layer technetium perovskite
Sr$_2$TcO$_4$. The calculated in-plane exchange interactions are large, which
establishes Sr$_2$TcO$_4$  as a strong 2D magnet.
We predict the long range AFM order to occur (despite the complete
frustration of the inter-plane exchange) due to the effective spin
anisotropy that 
originates in the dipole-dipole
interactions. We estimate the ordering temperature to be in the
500-600\,K range.

 We note that it would be interesting to dope this compound. Namely,
 Sr$_2$TcO$_4$ is according to our predictions a strong
 antiferromagnet, but the magnetic order must disappear upon doping
 with electrons, as Sr$_2$RuO$_4$ is known to be a (low-spin)
 paramagnet that becomes an unconventional superconductor below
 1.5\,K. 
 Similar would happen on doping with holes, as SrMoO$_3$ is
 paramagnetic, too.  In cuprates, strong magnetic interactions in a
 doped compound in which magnetic order disappears lead to an unusual
 electronic state and high-temperature
 superconductivity. \cite{dagotto_94} From iron-based superconductors,
 we learned that the high-temperature superconductivity is possible
 also in a multi-orbital context. \cite{paglione10} In doped Tc
 compounds similar, or other interesting discoveries may be ahead.  In
 this respect, not only investigations on layered Tc compounds
 \cite{hartmann2012} but also the attempts of material synthesis of
 half-filled t$_{2g}$ shell perovskites \cite{hiley14,seki14} are
 extremely interesting.

\begin{acknowledgements}{We warmly thank  Antoine Georges for stimulating and helpful discussions. 
L.P. acknowledges computational resources provided by the Swedish
National Infrastructure for Computing (SNIC) at National Supercomputer
Centre (NSC) and PDC Centre for High Performance Computing (PDC-HPC).
M.A. is supported by the Austrian Science Fund FWF, projects F04103
and Y746. J.M and A.H.  acknowledge support of Slovenian research
agency under program P1-0044.  }
 \end{acknowledgements}

\appendix

\section{Calculation of magnetic interactions within DFT+DMFT}
\label{app:leonid}
In the case of Tc$^{4+}$ and other transition-metal ions with a ground-state (GS) multiplet
well separated from excited states, the Hubbard-I atomic Green's function (GF) at low temperature has a particularly simple form $G_\mathrm{at}(i\omega_n)=\sum_{\Gamma}G_{\Gamma}(i\omega_n)w_{\Gamma}$. 
The matrix elements $G^{\sigma,m m'}_{\Gamma}$ of the contribution due to the state ${\Gamma}$ of the GS multiplet are given (in the imaginary time domain) by $-\langle\Gamma|T[c_{\sigma m}(\tau)c^{\dagger}_{\sigma m'}(0)]|\Gamma\rangle$, where $c^{(\dagger)}_{\sigma m}$ are annihilation (creation) operators for the spin $\sigma$ and magnetic quantum number $m$, $T$ is the time-ordering operator. $w_{\Gamma}$ are the weights of states within the GS multiplet, their ``atomic" values at low temperature (and in absence of crystal-field splitting) are simply given by $1/N$, where $N$ is the GS multiplicity. 

However, one may consider $G_\mathrm{at}$ (and the corresponding self-energy
$\Sigma_\mathrm{at}$ related to $G_\mathrm{at}$ by the Dyson equation) as a 
  function of $w_{\Gamma}$. By writing the DFT+DMFT free energy in
the standard form \cite{Savrasov2004} as a functional of the local
self-energy and GF, introducing small fluctuations of
$w^{i(j)}_{\Gamma}$ with respect to their ``atomic" value on two
neighboring sites $i$($j$) and then computing the linear response of
the free energy due to the fluctuation $\delta w^{i}_{\Gamma} \delta
w^{j}_{\Gamma'}$ one obtains the corresponding matrix element of
$\langle \Gamma_i \Gamma'_j|\mathcal{H}_{le}|\Gamma_i
\Gamma'_j\rangle$ of a low-energy spin-orbital Hamiltonian
$\mathcal{H}_{le}$. In the present case of half-filled TM ions
(e.g. Tc$^{4+}$) the atomic states $\{\Gamma\}$ are in fact just
different eigenstates of the $S_z$ operator, and the resulting matrix
elements are those of the quantum Heisenberg Hamiltonian
Eq.~(\ref{eq:spin_hamiltonian0}), from which $J_{ij}$ are easily
obtained. Detailed description of the method will be given in a separate publication.\cite{pourovskii14u}

In DFT+HubI calculations of Tc perovskites within Hubbard-I we employed
the values of orbital on-site $U=2.3$\,eV and $J_H=0.3$\,eV, as
described in the text. For other compounds considered we have selected 
the values (in eV) of $U$, $J_H$ consistent with available literature
data: 3.5,0.6 for SrMnO$_3$ and Sr$_2$MnO$_4$; 9.1,0.7 for KNiF$_3$
and K$_2$NiF$_4$; 5.1,0.7 for Rb$_2$MnF$_4$.

%


\begin{thebibliography}{58}%
\makeatletter
\providecommand \@ifxundefined [1]{%
 \@ifx{#1\undefined}
}%
\providecommand \@ifnum [1]{%
 \ifnum #1\expandafter \@firstoftwo
 \else \expandafter \@secondoftwo
 \fi
}%
\providecommand \@ifx [1]{%
 \ifx #1\expandafter \@firstoftwo
 \else \expandafter \@secondoftwo
 \fi
}%
\providecommand \natexlab [1]{#1}%
\providecommand \enquote  [1]{``#1''}%
\providecommand \bibnamefont  [1]{#1}%
\providecommand \bibfnamefont [1]{#1}%
\providecommand \citenamefont [1]{#1}%
\providecommand \href@noop [0]{\@secondoftwo}%
\providecommand \href [0]{\begingroup \@sanitize@url \@href}%
\providecommand \@href[1]{\@@startlink{#1}\@@href}%
\providecommand \@@href[1]{\endgroup#1\@@endlink}%
\providecommand \@sanitize@url [0]{\catcode `\\12\catcode `\$12\catcode
  `\&12\catcode `\#12\catcode `\^12\catcode `\_12\catcode `\%12\relax}%
\providecommand \@@startlink[1]{}%
\providecommand \@@endlink[0]{}%
\providecommand \url  [0]{\begingroup\@sanitize@url \@url }%
\providecommand \@url [1]{\endgroup\@href {#1}{\urlprefix }}%
\providecommand \urlprefix  [0]{URL }%
\providecommand \Eprint [0]{\href }%
\providecommand \doibase [0]{http://dx.doi.org/}%
\providecommand \selectlanguage [0]{\@gobble}%
\providecommand \bibinfo  [0]{\@secondoftwo}%
\providecommand \bibfield  [0]{\@secondoftwo}%
\providecommand \translation [1]{[#1]}%
\providecommand \BibitemOpen [0]{}%
\providecommand \bibitemStop [0]{}%
\providecommand \bibitemNoStop [0]{.\EOS\space}%
\providecommand \EOS [0]{\spacefactor3000\relax}%
\providecommand \BibitemShut  [1]{\csname bibitem#1\endcsname}%
\let\auto@bib@innerbib\@empty
\bibitem [{\citenamefont {Rodriguez}\ \emph {et~al.}(2011)\citenamefont
  {Rodriguez}, \citenamefont {Poineau}, \citenamefont {Llobet}, \citenamefont
  {Kennedy}, \citenamefont {Avdeev}, \citenamefont {Thorogood}, \citenamefont
  {Carter}, \citenamefont {Seshadri}, \citenamefont {Singh},\ and\
  \citenamefont {Cheetham}}]{rodriguez11}%
  \BibitemOpen
  \bibfield  {author} {\bibinfo {author} {\bibfnamefont {E.~E.}\ \bibnamefont
  {Rodriguez}}, \bibinfo {author} {\bibfnamefont {F.}~\bibnamefont {Poineau}},
  \bibinfo {author} {\bibfnamefont {A.}~\bibnamefont {Llobet}}, \bibinfo
  {author} {\bibfnamefont {B.~J.}\ \bibnamefont {Kennedy}}, \bibinfo {author}
  {\bibfnamefont {M.}~\bibnamefont {Avdeev}}, \bibinfo {author} {\bibfnamefont
  {G.~J.}\ \bibnamefont {Thorogood}}, \bibinfo {author} {\bibfnamefont {M.~L.}\
  \bibnamefont {Carter}}, \bibinfo {author} {\bibfnamefont {R.}~\bibnamefont
  {Seshadri}}, \bibinfo {author} {\bibfnamefont {D.~J.}\ \bibnamefont {Singh}},
  \ and\ \bibinfo {author} {\bibfnamefont {A.~K.}\ \bibnamefont {Cheetham}},\
  }\href {\doibase 10.1103/PhysRevLett.106.067201} {\bibfield  {journal}
  {\bibinfo  {journal} {Phys. Rev. Lett.}\ }\textbf {\bibinfo {volume} {106}},\
  \bibinfo {pages} {067201} (\bibinfo {year} {2011})}\BibitemShut {NoStop}%
\bibitem [{\citenamefont {Avdeev}\ \emph {et~al.}(2011)\citenamefont {Avdeev},
  \citenamefont {Thorogood}, \citenamefont {Carter}, \citenamefont {Kennedy},
  \citenamefont {Ting}, \citenamefont {Singh},\ and\ \citenamefont
  {Wallwork}}]{avdeev11}%
  \BibitemOpen
  \bibfield  {author} {\bibinfo {author} {\bibfnamefont {M.}~\bibnamefont
  {Avdeev}}, \bibinfo {author} {\bibfnamefont {G.~J.}\ \bibnamefont
  {Thorogood}}, \bibinfo {author} {\bibfnamefont {M.~L.}\ \bibnamefont
  {Carter}}, \bibinfo {author} {\bibfnamefont {B.~J.}\ \bibnamefont {Kennedy}},
  \bibinfo {author} {\bibfnamefont {J.}~\bibnamefont {Ting}}, \bibinfo {author}
  {\bibfnamefont {D.~J.}\ \bibnamefont {Singh}}, \ and\ \bibinfo {author}
  {\bibfnamefont {K.~S.}\ \bibnamefont {Wallwork}},\ }\href {\doibase
  10.1021/ja109431t} {\bibfield  {journal} {\bibinfo  {journal} {J. Am. Chem.
  Soc.}\ }\textbf {\bibinfo {volume} {133}},\ \bibinfo {pages} {1654} (\bibinfo
  {year} {2011})}\BibitemShut {NoStop}%
\bibitem [{\citenamefont {de' Medici}\ \emph {et~al.}(2011)\citenamefont {de'
  Medici}, \citenamefont {Mravlje},\ and\ \citenamefont
  {Georges}}]{demedici11}%
  \BibitemOpen
  \bibfield  {author} {\bibinfo {author} {\bibfnamefont {L.}~\bibnamefont {de'
  Medici}}, \bibinfo {author} {\bibfnamefont {J.}~\bibnamefont {Mravlje}}, \
  and\ \bibinfo {author} {\bibfnamefont {A.}~\bibnamefont {Georges}},\ }\href
  {\doibase 10.1103/PhysRevLett.107.256401} {\bibfield  {journal} {\bibinfo
  {journal} {Phys. Rev. Lett.}\ }\textbf {\bibinfo {volume} {107}},\ \bibinfo
  {pages} {256401} (\bibinfo {year} {2011})}\BibitemShut {NoStop}%
\bibitem [{\citenamefont {Franchini}\ \emph {et~al.}(2011)\citenamefont
  {Franchini}, \citenamefont {Archer}, \citenamefont {He}, \citenamefont
  {Chen}, \citenamefont {Filippetti},\ and\ \citenamefont
  {Sanvito}}]{franchini11}%
  \BibitemOpen
  \bibfield  {author} {\bibinfo {author} {\bibfnamefont {C.}~\bibnamefont
  {Franchini}}, \bibinfo {author} {\bibfnamefont {T.}~\bibnamefont {Archer}},
  \bibinfo {author} {\bibfnamefont {J.}~\bibnamefont {He}}, \bibinfo {author}
  {\bibfnamefont {X.-Q.}\ \bibnamefont {Chen}}, \bibinfo {author}
  {\bibfnamefont {A.}~\bibnamefont {Filippetti}}, \ and\ \bibinfo {author}
  {\bibfnamefont {S.}~\bibnamefont {Sanvito}},\ }\href {\doibase
  10.1103/PhysRevB.83.220402} {\bibfield  {journal} {\bibinfo  {journal} {Phys.
  Rev. B}\ }\textbf {\bibinfo {volume} {83}},\ \bibinfo {pages} {220402}
  (\bibinfo {year} {2011})}\BibitemShut {NoStop}%
\bibitem [{\citenamefont {Mravlje}\ \emph {et~al.}(2012)\citenamefont
  {Mravlje}, \citenamefont {Aichhorn},\ and\ \citenamefont
  {Georges}}]{mravlje12}%
  \BibitemOpen
  \bibfield  {author} {\bibinfo {author} {\bibfnamefont {J.}~\bibnamefont
  {Mravlje}}, \bibinfo {author} {\bibfnamefont {M.}~\bibnamefont {Aichhorn}}, \
  and\ \bibinfo {author} {\bibfnamefont {A.}~\bibnamefont {Georges}},\ }\href
  {\doibase 10.1103/PhysRevLett.108.197202} {\bibfield  {journal} {\bibinfo
  {journal} {Phys. Rev. Lett.}\ }\textbf {\bibinfo {volume} {108}},\ \bibinfo
  {pages} {197202} (\bibinfo {year} {2012})}\BibitemShut {NoStop}%
\bibitem [{\citenamefont {Middey}\ \emph {et~al.}(2012)\citenamefont {Middey},
  \citenamefont {Nandy}, \citenamefont {Pandey}, \citenamefont {Mahadevan},\
  and\ \citenamefont {Sarma}}]{middey_sarma_12}%
  \BibitemOpen
  \bibfield  {author} {\bibinfo {author} {\bibfnamefont {S.}~\bibnamefont
  {Middey}}, \bibinfo {author} {\bibfnamefont {A.~K.}\ \bibnamefont {Nandy}},
  \bibinfo {author} {\bibfnamefont {S.~K.}\ \bibnamefont {Pandey}}, \bibinfo
  {author} {\bibfnamefont {P.}~\bibnamefont {Mahadevan}}, \ and\ \bibinfo
  {author} {\bibfnamefont {D.~D.}\ \bibnamefont {Sarma}},\ }\href {\doibase
  10.1103/PhysRevB.86.104406} {\bibfield  {journal} {\bibinfo  {journal} {Phys.
  Rev. B}\ }\textbf {\bibinfo {volume} {86}},\ \bibinfo {pages} {104406}
  (\bibinfo {year} {2012})}\BibitemShut {NoStop}%
\bibitem [{\citenamefont {Wang}\ \emph {et~al.}(2012)\citenamefont {Wang},
  \citenamefont {Li}, \citenamefont {Liu}, \citenamefont {Zhang},\ and\
  \citenamefont {Yang}}]{wang12}%
  \BibitemOpen
  \bibfield  {author} {\bibinfo {author} {\bibfnamefont {G.}~\bibnamefont
  {Wang}}, \bibinfo {author} {\bibfnamefont {L.}~\bibnamefont {Li}}, \bibinfo
  {author} {\bibfnamefont {C.}~\bibnamefont {Liu}}, \bibinfo {author}
  {\bibfnamefont {M.}~\bibnamefont {Zhang}}, \ and\ \bibinfo {author}
  {\bibfnamefont {Z.}~\bibnamefont {Yang}},\ }\href@noop {} {\bibfield
  {journal} {\bibinfo  {journal} {Phys. Lett. A}\ }\textbf {\bibinfo {volume}
  {376}},\ \bibinfo {pages} {3313} (\bibinfo {year} {2012})}\BibitemShut
  {NoStop}%
\bibitem [{\citenamefont {Dai}\ and\ \citenamefont {Ma}(2014)}]{dai14}%
  \BibitemOpen
  \bibfield  {author} {\bibinfo {author} {\bibfnamefont {C.-M.}\ \bibnamefont
  {Dai}}\ and\ \bibinfo {author} {\bibfnamefont {C.-L.}\ \bibnamefont {Ma}},\
  }\href@noop {} {\bibfield  {journal} {\bibinfo  {journal} {Mod. Phys. Lett.}\
  }\textbf {\bibinfo {volume} {28}},\ \bibinfo {pages} {1450049} (\bibinfo
  {year} {2014})}\BibitemShut {NoStop}%
\bibitem [{\citenamefont {Georges}\ \emph {et~al.}(2013)\citenamefont
  {Georges}, \citenamefont {de'Medici},\ and\ \citenamefont
  {Mravlje}}]{georges_annrev13}%
  \BibitemOpen
  \bibfield  {author} {\bibinfo {author} {\bibfnamefont {A.}~\bibnamefont
  {Georges}}, \bibinfo {author} {\bibfnamefont {L.}~\bibnamefont {de'Medici}},
  \ and\ \bibinfo {author} {\bibfnamefont {J.}~\bibnamefont {Mravlje}},\
  }\href@noop {} {\bibfield  {journal} {\bibinfo  {journal} {Annu. Rev.
  Condens. Matter Phys. 4, 137 (2013)}\ }\textbf {\bibinfo {volume} {4}},\
  \bibinfo {pages} {137} (\bibinfo {year} {2013})}\BibitemShut {NoStop}%
\bibitem [{\citenamefont {Pies}\ and\ \citenamefont {Weiss}()}]{Goodenough70}%
  \BibitemOpen
  \bibfield  {author} {\bibinfo {author} {\bibfnamefont {W.}~\bibnamefont
  {Pies}}\ and\ \bibinfo {author} {\bibfnamefont {A.}~\bibnamefont {Weiss}},\
  }\enquote {\bibinfo {title} {Landolt-bornstein - group iii condensed
  matter},}\ Chap.\ \bibinfo {chapter} {f2694, XX.2.1 Simple oxo-compounds of
  technetium (oxotechnetates)}\BibitemShut {NoStop}%
\bibitem [{\citenamefont {Hartmann}\ \emph {et~al.}(2012)\citenamefont
  {Hartmann}, \citenamefont {Alaniz},\ and\ \citenamefont
  {Antonio}}]{hartmann2012}%
  \BibitemOpen
  \bibfield  {author} {\bibinfo {author} {\bibfnamefont {T.}~\bibnamefont
  {Hartmann}}, \bibinfo {author} {\bibfnamefont {A.~J.}\ \bibnamefont
  {Alaniz}}, \ and\ \bibinfo {author} {\bibfnamefont {D.~J.}\ \bibnamefont
  {Antonio}},\ }\href {\doibase 10.1016/j.proche.2012.10.095} {\bibfield
  {journal} {\bibinfo  {journal} {Proc. Chem.}\ }\textbf {\bibinfo {volume}
  {7}},\ \bibinfo {pages} {622} (\bibinfo {year} {2012})}\BibitemShut {NoStop}%
\bibitem [{\citenamefont {Mermin}\ and\ \citenamefont
  {Wagner}(1966)}]{mermin1966}%
  \BibitemOpen
  \bibfield  {author} {\bibinfo {author} {\bibfnamefont {N.~D.}\ \bibnamefont
  {Mermin}}\ and\ \bibinfo {author} {\bibfnamefont {H.}~\bibnamefont
  {Wagner}},\ }\href
  {http://journals.aps.org/prl/abstract/10.1103/PhysRevLett.17.1133} {\bibfield
   {journal} {\bibinfo  {journal} {Phys. Rev. Lett.}\ }\textbf {\bibinfo
  {volume} {17}},\ \bibinfo {pages} {1133} (\bibinfo {year}
  {1966})}\BibitemShut {NoStop}%
\bibitem [{\citenamefont {Chakravarty}\ \emph {et~al.}(1989)\citenamefont
  {Chakravarty}, \citenamefont {Halperin},\ and\ \citenamefont
  {Nelson}}]{chakravarty89}%
  \BibitemOpen
  \bibfield  {author} {\bibinfo {author} {\bibfnamefont {S.}~\bibnamefont
  {Chakravarty}}, \bibinfo {author} {\bibfnamefont {B.~I.}\ \bibnamefont
  {Halperin}}, \ and\ \bibinfo {author} {\bibfnamefont {D.~R.}\ \bibnamefont
  {Nelson}},\ }\href {\doibase 10.1103/PhysRevB.39.2344} {\bibfield  {journal}
  {\bibinfo  {journal} {Phys. Rev. B}\ }\textbf {\bibinfo {volume} {39}},\
  \bibinfo {pages} {2344} (\bibinfo {year} {1989})}\BibitemShut {NoStop}%
\bibitem [{\citenamefont {Scalapino}\ \emph {et~al.}(1975)\citenamefont
  {Scalapino}, \citenamefont {Imry},\ and\ \citenamefont
  {Pincus}}]{scalapino75}%
  \BibitemOpen
  \bibfield  {author} {\bibinfo {author} {\bibfnamefont {D.~J.}\ \bibnamefont
  {Scalapino}}, \bibinfo {author} {\bibfnamefont {Y.}~\bibnamefont {Imry}}, \
  and\ \bibinfo {author} {\bibfnamefont {P.}~\bibnamefont {Pincus}},\ }\href
  {\doibase 10.1103/PhysRevB.11.2042} {\bibfield  {journal} {\bibinfo
  {journal} {Phys. Rev. B}\ }\textbf {\bibinfo {volume} {11}},\ \bibinfo
  {pages} {2042} (\bibinfo {year} {1975})}\BibitemShut {NoStop}%
\bibitem [{\citenamefont {Schulz}(1996)}]{schulz96}%
  \BibitemOpen
  \bibfield  {author} {\bibinfo {author} {\bibfnamefont {H.~J.}\ \bibnamefont
  {Schulz}},\ }\href {\doibase 10.1103/PhysRevLett.77.2790} {\bibfield
  {journal} {\bibinfo  {journal} {Phys. Rev. Lett.}\ }\textbf {\bibinfo
  {volume} {77}},\ \bibinfo {pages} {2790} (\bibinfo {year}
  {1996})}\BibitemShut {NoStop}%
\bibitem [{\citenamefont {Yasuda}\ \emph {et~al.}(2005)\citenamefont {Yasuda},
  \citenamefont {Todo}, \citenamefont {Hukushima}, \citenamefont {Alet},
  \citenamefont {Keller}, \citenamefont {Troyer},\ and\ \citenamefont
  {Takayama}}]{yasuda05}%
  \BibitemOpen
  \bibfield  {author} {\bibinfo {author} {\bibfnamefont {C.}~\bibnamefont
  {Yasuda}}, \bibinfo {author} {\bibfnamefont {S.}~\bibnamefont {Todo}},
  \bibinfo {author} {\bibfnamefont {K.}~\bibnamefont {Hukushima}}, \bibinfo
  {author} {\bibfnamefont {F.}~\bibnamefont {Alet}}, \bibinfo {author}
  {\bibfnamefont {M.}~\bibnamefont {Keller}}, \bibinfo {author} {\bibfnamefont
  {M.}~\bibnamefont {Troyer}}, \ and\ \bibinfo {author} {\bibfnamefont
  {H.}~\bibnamefont {Takayama}},\ }\href
  {http://dx.doi.org/10.1103/PhysRevLett.94.217201} {\bibfield  {journal}
  {\bibinfo  {journal} {Phys. Rev. Lett.}\ }\textbf {\bibinfo {volume} {94}},\
  \bibinfo {pages} {217201} (\bibinfo {year} {2005})}\BibitemShut {NoStop}%
\bibitem [{\citenamefont {Lines}(1967)}]{lines67}%
  \BibitemOpen
  \bibfield  {author} {\bibinfo {author} {\bibfnamefont {M.}~\bibnamefont
  {Lines}},\ }\href {\doibase 10.1103/physrev.164.736} {\bibfield  {journal}
  {\bibinfo  {journal} {Phys. Rev.}\ }\textbf {\bibinfo {volume} {164}},\
  \bibinfo {pages} {736} (\bibinfo {year} {1967})}\BibitemShut {NoStop}%
\bibitem [{\citenamefont {Zhou}\ \emph {et~al.}(2007)\citenamefont {Zhou},
  \citenamefont {Landau},\ and\ \citenamefont {Schulthess}}]{zhou07}%
  \BibitemOpen
  \bibfield  {author} {\bibinfo {author} {\bibfnamefont {C.}~\bibnamefont
  {Zhou}}, \bibinfo {author} {\bibfnamefont {D.}~\bibnamefont {Landau}}, \ and\
  \bibinfo {author} {\bibfnamefont {T.}~\bibnamefont {Schulthess}},\ }\href
  {http://dx.doi.org/10.1103/PhysRevB.76.024433} {\bibfield  {journal}
  {\bibinfo  {journal} {Phys. Rev. B}\ }\textbf {\bibinfo {volume} {76}},\
  \bibinfo {pages} {024433} (\bibinfo {year} {2007})}\BibitemShut {NoStop}%
\bibitem [{\citenamefont {Cowley}\ \emph {et~al.}(1977)\citenamefont {Cowley},
  \citenamefont {Shirane}, \citenamefont {Birgeneau},\ and\ \citenamefont
  {Guggenheim}}]{cowley_birgenau_1977}%
  \BibitemOpen
  \bibfield  {author} {\bibinfo {author} {\bibfnamefont {R.~A.}\ \bibnamefont
  {Cowley}}, \bibinfo {author} {\bibfnamefont {G.}~\bibnamefont {Shirane}},
  \bibinfo {author} {\bibfnamefont {R.~J.}\ \bibnamefont {Birgeneau}}, \ and\
  \bibinfo {author} {\bibfnamefont {H.~J.}\ \bibnamefont {Guggenheim}},\ }\href
  {\doibase 10.1103/PhysRevB.15.4292} {\bibfield  {journal} {\bibinfo
  {journal} {Phys. Rev. B}\ }\textbf {\bibinfo {volume} {15}},\ \bibinfo
  {pages} {4292} (\bibinfo {year} {1977})}\BibitemShut {NoStop}%
\bibitem [{\citenamefont {Christianson}\ \emph {et~al.}(2001)\citenamefont
  {Christianson}, \citenamefont {Leheny}, \citenamefont {Birgeneau},\ and\
  \citenamefont {Erwin}}]{christianson_birgenau_2001}%
  \BibitemOpen
  \bibfield  {author} {\bibinfo {author} {\bibfnamefont {R.~J.}\ \bibnamefont
  {Christianson}}, \bibinfo {author} {\bibfnamefont {R.~L.}\ \bibnamefont
  {Leheny}}, \bibinfo {author} {\bibfnamefont {R.~J.}\ \bibnamefont
  {Birgeneau}}, \ and\ \bibinfo {author} {\bibfnamefont {R.~W.}\ \bibnamefont
  {Erwin}},\ }\href {\doibase 10.1103/PhysRevB.63.140401} {\bibfield  {journal}
  {\bibinfo  {journal} {Phys. Rev. B}\ }\textbf {\bibinfo {volume} {63}},\
  \bibinfo {pages} {140401} (\bibinfo {year} {2001})}\BibitemShut {NoStop}%
\bibitem [{\citenamefont {Landau}\ and\ \citenamefont
  {Binder}(2009)}]{landau_binder}%
  \BibitemOpen
  \bibfield  {author} {\bibinfo {author} {\bibfnamefont {D.~P.}\ \bibnamefont
  {Landau}}\ and\ \bibinfo {author} {\bibfnamefont {K.}~\bibnamefont
  {Binder}},\ }\href@noop {} {\emph {\bibinfo {title} {A guide to Monte Carlo
  simulations in statistical physics}}}\ (\bibinfo  {publisher} {Cambridge
  university press},\ \bibinfo {year} {2009})\BibitemShut {NoStop}%
\bibitem [{\citenamefont {R{\"o}{\ss}ler}(1999)}]{Roesler1999}%
  \BibitemOpen
  \bibfield  {author} {\bibinfo {author} {\bibfnamefont {U.}~\bibnamefont
  {R{\"o}{\ss}ler}},\ }\href
  {http://journals.aps.org/prb/abstract/10.1103/PhysRevB.59.13577} {\bibfield
  {journal} {\bibinfo  {journal} {Phys. Rev. B}\ }\textbf {\bibinfo {volume}
  {59}},\ \bibinfo {pages} {13577} (\bibinfo {year} {1999})}\BibitemShut
  {NoStop}%
\bibitem [{\citenamefont {Blaha}\ \emph {et~al.}(2001)\citenamefont {Blaha},
  \citenamefont {Schwarz}, \citenamefont {Madsen}, \citenamefont {Kvasnicka},\
  and\ \citenamefont {Luitz}}]{Wien2k}%
  \BibitemOpen
  \bibfield  {author} {\bibinfo {author} {\bibfnamefont {P.}~\bibnamefont
  {Blaha}}, \bibinfo {author} {\bibfnamefont {K.}~\bibnamefont {Schwarz}},
  \bibinfo {author} {\bibfnamefont {G.}~\bibnamefont {Madsen}}, \bibinfo
  {author} {\bibfnamefont {D.}~\bibnamefont {Kvasnicka}}, \ and\ \bibinfo
  {author} {\bibfnamefont {J.}~\bibnamefont {Luitz}},\ }\href@noop {} {\emph
  {\bibinfo {title} {WIEN2k, An augmented Plane Wave + Local Orbitals Program
  for Calculating Crystal Properties}}}\ (\bibinfo  {publisher} {Techn.
  Universitat Wien, Austria, ISBN 3-9501031-1-2.},\ \bibinfo {year}
  {2001})\BibitemShut {NoStop}%
\bibitem [{Note1()}]{Note1}%
  \BibitemOpen
  \bibinfo {note} {DFT+U calculations were carried out using the implementation
  of the Wien2k package within GGA approximation. The correlated orbitals used
  in this implementation are different from those employed in the TRIQS
  package, hence, we employed slightly different values of U in our DFT+DMFT
  and DFT+U calculations of Sr$_2$TcO$_4$}\BibitemShut {NoStop}%
\bibitem [{\citenamefont {Anisimov}\ \emph {et~al.}(1993)\citenamefont
  {Anisimov}, \citenamefont {Solovyev}, \citenamefont {Korotin}, \citenamefont
  {Czy\ifmmode~\dot{z}\else \.{z}\fi{}yk},\ and\ \citenamefont
  {Sawatzky}}]{Anisimov1993}%
  \BibitemOpen
  \bibfield  {author} {\bibinfo {author} {\bibfnamefont {V.~I.}\ \bibnamefont
  {Anisimov}}, \bibinfo {author} {\bibfnamefont {I.~V.}\ \bibnamefont
  {Solovyev}}, \bibinfo {author} {\bibfnamefont {M.~A.}\ \bibnamefont
  {Korotin}}, \bibinfo {author} {\bibfnamefont {M.~T.}\ \bibnamefont
  {Czy\ifmmode~\dot{z}\else \.{z}\fi{}yk}}, \ and\ \bibinfo {author}
  {\bibfnamefont {G.~A.}\ \bibnamefont {Sawatzky}},\ }\href {\doibase
  10.1103/PhysRevB.48.16929} {\bibfield  {journal} {\bibinfo  {journal} {Phys.
  Rev. B}\ }\textbf {\bibinfo {volume} {48}},\ \bibinfo {pages} {16929}
  (\bibinfo {year} {1993})}\BibitemShut {NoStop}%
\bibitem [{\citenamefont {Perdew}\ \emph {et~al.}(1996)\citenamefont {Perdew},
  \citenamefont {Burke},\ and\ \citenamefont {Ernzerhof}}]{PBE96}%
  \BibitemOpen
  \bibfield  {author} {\bibinfo {author} {\bibfnamefont {J.~P.}\ \bibnamefont
  {Perdew}}, \bibinfo {author} {\bibfnamefont {K.}~\bibnamefont {Burke}}, \
  and\ \bibinfo {author} {\bibfnamefont {M.}~\bibnamefont {Ernzerhof}},\ }\href
  {\doibase 10.1103/PhysRevLett.77.3865} {\bibfield  {journal} {\bibinfo
  {journal} {Phys. Rev. Lett.}\ }\textbf {\bibinfo {volume} {77}},\ \bibinfo
  {pages} {3865} (\bibinfo {year} {1996})}\BibitemShut {NoStop}%
\bibitem [{\citenamefont {Ferrero}\ and\ \citenamefont {Parcollet}()}]{triqs}%
  \BibitemOpen
  \bibfield  {author} {\bibinfo {author} {\bibfnamefont {M.}~\bibnamefont
  {Ferrero}}\ and\ \bibinfo {author} {\bibfnamefont {O.}~\bibnamefont
  {Parcollet}},\ }\href@noop {} {\enquote {\bibinfo {title} {Triqs: a toolkit
  for research in interacting quantum systems},}\ }\bibinfo {note}
  {Http://ipht.cea.fr/triqs}\BibitemShut {NoStop}%
\bibitem [{\citenamefont {Aichhorn}\ \emph {et~al.}(2009)\citenamefont
  {Aichhorn} \emph {et~al.}}]{aichhorn09}%
  \BibitemOpen
  \bibfield  {author} {\bibinfo {author} {\bibfnamefont {M.}~\bibnamefont
  {Aichhorn}} \emph {et~al.},\ }\href {\doibase 10.1103/PhysRevB.80.085101}
  {\bibfield  {journal} {\bibinfo  {journal} {Phys. Rev. B}\ }\textbf {\bibinfo
  {volume} {80}},\ \bibinfo {pages} {085101} (\bibinfo {year}
  {2009})}\BibitemShut {NoStop}%
\bibitem [{\citenamefont {Aichhorn}\ \emph {et~al.}(2011)\citenamefont
  {Aichhorn}, \citenamefont {Pourovskii},\ and\ \citenamefont
  {Georges}}]{aichhorn11}%
  \BibitemOpen
  \bibfield  {author} {\bibinfo {author} {\bibfnamefont {M.}~\bibnamefont
  {Aichhorn}}, \bibinfo {author} {\bibfnamefont {L.}~\bibnamefont
  {Pourovskii}}, \ and\ \bibinfo {author} {\bibfnamefont {A.}~\bibnamefont
  {Georges}},\ }\href {\doibase 10.1103/PhysRevB.84.054529} {\bibfield
  {journal} {\bibinfo  {journal} {Phys. Rev. B}\ }\textbf {\bibinfo {volume}
  {84}},\ \bibinfo {pages} {054529} (\bibinfo {year} {2011})}\BibitemShut
  {NoStop}%
\bibitem [{\citenamefont {Gull}\ \emph {et~al.}(2011)\citenamefont {Gull},
  \citenamefont {Millis}, \citenamefont {Lichtenstein}, \citenamefont
  {Rubtsov}, \citenamefont {Troyer},\ and\ \citenamefont {Werner}}]{gull11}%
  \BibitemOpen
  \bibfield  {author} {\bibinfo {author} {\bibfnamefont {E.}~\bibnamefont
  {Gull}}, \bibinfo {author} {\bibfnamefont {A.~J.}\ \bibnamefont {Millis}},
  \bibinfo {author} {\bibfnamefont {A.~I.}\ \bibnamefont {Lichtenstein}},
  \bibinfo {author} {\bibfnamefont {A.~N.}\ \bibnamefont {Rubtsov}}, \bibinfo
  {author} {\bibfnamefont {M.}~\bibnamefont {Troyer}}, \ and\ \bibinfo {author}
  {\bibfnamefont {P.}~\bibnamefont {Werner}},\ }\href {\doibase
  10.1103/RevModPhys.83.349} {\bibfield  {journal} {\bibinfo  {journal} {Rev.
  Mod. Phys.}\ }\textbf {\bibinfo {volume} {83}},\ \bibinfo {pages} {349}
  (\bibinfo {year} {2011})}\BibitemShut {NoStop}%
\bibitem [{\citenamefont {Werner}\ and\ \citenamefont
  {Millis}(2006)}]{werner2}%
  \BibitemOpen
  \bibfield  {author} {\bibinfo {author} {\bibfnamefont {P.}~\bibnamefont
  {Werner}}\ and\ \bibinfo {author} {\bibfnamefont {A.~J.}\ \bibnamefont
  {Millis}},\ }\href {\doibase 10.1103/PhysRevB.74.155107} {\bibfield
  {journal} {\bibinfo  {journal} {Phys. Rev. B}\ }\textbf {\bibinfo {volume}
  {74}},\ \bibinfo {pages} {155107} (\bibinfo {year} {2006})}\BibitemShut
  {NoStop}%
\bibitem [{\citenamefont {Boehnke}\ \emph {et~al.}(2011)\citenamefont
  {Boehnke}, \citenamefont {Hafermann}, \citenamefont {Ferrero}, \citenamefont
  {Lechermann},\ and\ \citenamefont {Parcollet}}]{boehnke11}%
  \BibitemOpen
  \bibfield  {author} {\bibinfo {author} {\bibfnamefont {L.}~\bibnamefont
  {Boehnke}}, \bibinfo {author} {\bibfnamefont {H.}~\bibnamefont {Hafermann}},
  \bibinfo {author} {\bibfnamefont {M.}~\bibnamefont {Ferrero}}, \bibinfo
  {author} {\bibfnamefont {F.}~\bibnamefont {Lechermann}}, \ and\ \bibinfo
  {author} {\bibfnamefont {O.}~\bibnamefont {Parcollet}},\ }\href {\doibase
  10.1103/PhysRevB.84.075145} {\bibfield  {journal} {\bibinfo  {journal} {Phys.
  Rev. B}\ }\textbf {\bibinfo {volume} {84}},\ \bibinfo {pages} {075145}
  (\bibinfo {year} {2011})}\BibitemShut {NoStop}%
\bibitem [{\citenamefont {Parragh}\ \emph {et~al.}(2012)\citenamefont
  {Parragh}, \citenamefont {Toschi}, \citenamefont {Held},\ and\ \citenamefont
  {Sangiovanni}}]{parragh12}%
  \BibitemOpen
  \bibfield  {author} {\bibinfo {author} {\bibfnamefont {N.}~\bibnamefont
  {Parragh}}, \bibinfo {author} {\bibfnamefont {A.}~\bibnamefont {Toschi}},
  \bibinfo {author} {\bibfnamefont {K.}~\bibnamefont {Held}}, \ and\ \bibinfo
  {author} {\bibfnamefont {G.}~\bibnamefont {Sangiovanni}},\ }\href {\doibase
  10.1103/PhysRevB.86.155158} {\bibfield  {journal} {\bibinfo  {journal} {Phys.
  Rev. B}\ }\textbf {\bibinfo {volume} {86}},\ \bibinfo {pages} {155158}
  (\bibinfo {year} {2012})}\BibitemShut {NoStop}%
\bibitem [{\citenamefont {Hubbard}(1963)}]{hubbard_1}%
  \BibitemOpen
  \bibfield  {author} {\bibinfo {author} {\bibfnamefont {J.}~\bibnamefont
  {Hubbard}},\ }\href@noop {} {\bibfield  {journal} {\bibinfo  {journal} {Proc.
  Roy. Soc. (London)}\ }\textbf {\bibinfo {volume} {A 276}},\ \bibinfo {pages}
  {238} (\bibinfo {year} {1963})}\BibitemShut {NoStop}%
\bibitem [{\citenamefont {Liechtenstein}\ \emph {et~al.}(1987)\citenamefont
  {Liechtenstein}, \citenamefont {Katsnelson}, \citenamefont {Antropov},\ and\
  \citenamefont {Gubanov}}]{Liechtenstein1987}%
  \BibitemOpen
  \bibfield  {author} {\bibinfo {author} {\bibfnamefont {A.}~\bibnamefont
  {Liechtenstein}}, \bibinfo {author} {\bibfnamefont {M.}~\bibnamefont
  {Katsnelson}}, \bibinfo {author} {\bibfnamefont {V.}~\bibnamefont
  {Antropov}}, \ and\ \bibinfo {author} {\bibfnamefont {V.}~\bibnamefont
  {Gubanov}},\ }\href@noop {} {\bibfield  {journal} {\bibinfo  {journal} {J.
  Magn. Magn. Mater.}\ }\textbf {\bibinfo {volume} {67}},\ \bibinfo {pages} {65
  } (\bibinfo {year} {1987})}\BibitemShut {NoStop}%
\bibitem [{\citenamefont {Bruno}(2003)}]{Bruno2003}%
  \BibitemOpen
  \bibfield  {author} {\bibinfo {author} {\bibfnamefont {P.}~\bibnamefont
  {Bruno}},\ }\href@noop {} {\bibfield  {journal} {\bibinfo  {journal} {Phys.
  Rev. Lett.}\ }\textbf {\bibinfo {volume} {90}},\ \bibinfo {pages} {087205}
  (\bibinfo {year} {2003})}\BibitemShut {NoStop}%
\bibitem [{\citenamefont {Ruban}\ \emph {et~al.}(2004)\citenamefont {Ruban},
  \citenamefont {Shallcross}, \citenamefont {Simak},\ and\ \citenamefont
  {Skriver}}]{Ruban2004}%
  \BibitemOpen
  \bibfield  {author} {\bibinfo {author} {\bibfnamefont {A.~V.}\ \bibnamefont
  {Ruban}}, \bibinfo {author} {\bibfnamefont {S.}~\bibnamefont {Shallcross}},
  \bibinfo {author} {\bibfnamefont {S.~I.}\ \bibnamefont {Simak}}, \ and\
  \bibinfo {author} {\bibfnamefont {H.~L.}\ \bibnamefont {Skriver}},\
  }\href@noop {} {\bibfield  {journal} {\bibinfo  {journal} {Phys. Rev. B}\
  }\textbf {\bibinfo {volume} {70}},\ \bibinfo {pages} {125115} (\bibinfo
  {year} {2004})}\BibitemShut {NoStop}%
\bibitem [{\citenamefont {Katsnelson}\ and\ \citenamefont
  {Lichtenstein}(2000)}]{Katsnelson2000}%
  \BibitemOpen
  \bibfield  {author} {\bibinfo {author} {\bibfnamefont {M.~I.}\ \bibnamefont
  {Katsnelson}}\ and\ \bibinfo {author} {\bibfnamefont {A.~I.}\ \bibnamefont
  {Lichtenstein}},\ }\href@noop {} {\bibfield  {journal} {\bibinfo  {journal}
  {Phys. Rev. B}\ }\textbf {\bibinfo {volume} {61}},\ \bibinfo {pages} {8906}
  (\bibinfo {year} {2000})}\BibitemShut {NoStop}%
\bibitem [{\citenamefont {Wan}\ \emph {et~al.}(2006)\citenamefont {Wan},
  \citenamefont {Yin},\ and\ \citenamefont {Savrasov}}]{Wan2006}%
  \BibitemOpen
  \bibfield  {author} {\bibinfo {author} {\bibfnamefont {X.}~\bibnamefont
  {Wan}}, \bibinfo {author} {\bibfnamefont {Q.}~\bibnamefont {Yin}}, \ and\
  \bibinfo {author} {\bibfnamefont {S.~Y.}\ \bibnamefont {Savrasov}},\
  }\href@noop {} {\bibfield  {journal} {\bibinfo  {journal} {Phys. Rev. Lett.}\
  }\textbf {\bibinfo {volume} {97}},\ \bibinfo {pages} {266403} (\bibinfo
  {year} {2006})}\BibitemShut {NoStop}%
\bibitem [{\citenamefont {Rushbrooke}\ and\ \citenamefont
  {Wood}(1963)}]{Rushbrooke1963}%
  \BibitemOpen
  \bibfield  {author} {\bibinfo {author} {\bibfnamefont {G.}~\bibnamefont
  {Rushbrooke}}\ and\ \bibinfo {author} {\bibfnamefont {P.}~\bibnamefont
  {Wood}},\ }\href {\doibase 10.1080/00268976300100461} {\bibfield  {journal}
  {\bibinfo  {journal} {Mol. Phys.}\ }\textbf {\bibinfo {volume} {6}},\
  \bibinfo {pages} {409–421} (\bibinfo {year} {1963})}\BibitemShut {NoStop}%
\bibitem [{\citenamefont {Oitmaa}\ and\ \citenamefont
  {Zheng}(2004)}]{oitmaa04}%
  \BibitemOpen
  \bibfield  {author} {\bibinfo {author} {\bibfnamefont {J.}~\bibnamefont
  {Oitmaa}}\ and\ \bibinfo {author} {\bibfnamefont {W.}~\bibnamefont {Zheng}},\
  }\href {\doibase 10.1088/0953-8984/16/47/016} {\bibfield  {journal} {\bibinfo
   {journal} {J. Phys. Condens. Matter}\ }\textbf {\bibinfo {volume} {16}},\
  \bibinfo {pages} {8653} (\bibinfo {year} {2004})}\BibitemShut {NoStop}%
\bibitem [{\citenamefont {Martin}\ and\ \citenamefont
  {Illas}(1997)}]{Martin1997}%
  \BibitemOpen
  \bibfield  {author} {\bibinfo {author} {\bibfnamefont {R.}~\bibnamefont
  {Martin}}\ and\ \bibinfo {author} {\bibfnamefont {F.}~\bibnamefont {Illas}},\
  }\href {\doibase 10.1103/physrevlett.79.1539} {\bibfield  {journal} {\bibinfo
   {journal} {Phys. Rev. Lett.}\ }\textbf {\bibinfo {volume} {79}},\ \bibinfo
  {pages} {1539} (\bibinfo {year} {1997})}\BibitemShut {NoStop}%
\bibitem [{\citenamefont {Archer}\ \emph {et~al.}(2011)\citenamefont {Archer},
  \citenamefont {Pemmaraju}, \citenamefont {Sanvito}, \citenamefont
  {Franchini}, \citenamefont {He}, \citenamefont {Filippetti}, \citenamefont
  {Delugas}, \citenamefont {Puggioni}, \citenamefont {Fiorentini},
  \citenamefont {Tiwari},\ and\ \citenamefont {Majumdar}}]{archer11}%
  \BibitemOpen
  \bibfield  {author} {\bibinfo {author} {\bibfnamefont {T.}~\bibnamefont
  {Archer}}, \bibinfo {author} {\bibfnamefont {C.~D.}\ \bibnamefont
  {Pemmaraju}}, \bibinfo {author} {\bibfnamefont {S.}~\bibnamefont {Sanvito}},
  \bibinfo {author} {\bibfnamefont {C.}~\bibnamefont {Franchini}}, \bibinfo
  {author} {\bibfnamefont {J.}~\bibnamefont {He}}, \bibinfo {author}
  {\bibfnamefont {A.}~\bibnamefont {Filippetti}}, \bibinfo {author}
  {\bibfnamefont {P.}~\bibnamefont {Delugas}}, \bibinfo {author} {\bibfnamefont
  {D.}~\bibnamefont {Puggioni}}, \bibinfo {author} {\bibfnamefont
  {V.}~\bibnamefont {Fiorentini}}, \bibinfo {author} {\bibfnamefont
  {R.}~\bibnamefont {Tiwari}}, \ and\ \bibinfo {author} {\bibfnamefont
  {P.}~\bibnamefont {Majumdar}},\ }\href {\doibase 10.1103/PhysRevB.84.115114}
  {\bibfield  {journal} {\bibinfo  {journal} {Phys. Rev. B}\ }\textbf {\bibinfo
  {volume} {84}},\ \bibinfo {pages} {115114} (\bibinfo {year}
  {2011})}\BibitemShut {NoStop}%
\bibitem [{\citenamefont {{Beach}}()}]{Beach2004}%
  \BibitemOpen
  \bibfield  {author} {\bibinfo {author} {\bibfnamefont {K.~S.~D.}\
  \bibnamefont {{Beach}}},\ }\href@noop {} {\enquote {\bibinfo {title}
  {{Identifying the maximum entropy method as a special limit of stochastic
  analytic continuation}},}\ }\bibinfo {note} {Cond-mat/0403055}\BibitemShut
  {NoStop}%
\bibitem [{\citenamefont {Mravlje}\ \emph {et~al.}(2011)\citenamefont
  {Mravlje}, \citenamefont {Aichhorn}, \citenamefont {Miyake}, \citenamefont
  {Haule}, \citenamefont {Kotliar},\ and\ \citenamefont {Georges}}]{mravlje11}%
  \BibitemOpen
  \bibfield  {author} {\bibinfo {author} {\bibfnamefont {J.}~\bibnamefont
  {Mravlje}}, \bibinfo {author} {\bibfnamefont {M.}~\bibnamefont {Aichhorn}},
  \bibinfo {author} {\bibfnamefont {T.}~\bibnamefont {Miyake}}, \bibinfo
  {author} {\bibfnamefont {K.}~\bibnamefont {Haule}}, \bibinfo {author}
  {\bibfnamefont {G.}~\bibnamefont {Kotliar}}, \ and\ \bibinfo {author}
  {\bibfnamefont {A.}~\bibnamefont {Georges}},\ }\href@noop {} {\bibfield
  {journal} {\bibinfo  {journal} {Phys. Rev. Lett.}\ }\textbf {\bibinfo
  {volume} {106}},\ \bibinfo {pages} {096401} (\bibinfo {year}
  {2011})}\BibitemShut {NoStop}%
\bibitem [{\citenamefont {de~Jongh}\ and\ \citenamefont
  {Miedema}(1974)}]{jongh74}%
  \BibitemOpen
  \bibfield  {author} {\bibinfo {author} {\bibfnamefont {L.}~\bibnamefont
  {de~Jongh}}\ and\ \bibinfo {author} {\bibfnamefont {A.}~\bibnamefont
  {Miedema}},\ }\href {\doibase 10.1080/00018739700101558} {\bibfield
  {journal} {\bibinfo  {journal} {Adv. Phys.}\ }\textbf {\bibinfo {volume}
  {23}},\ \bibinfo {pages} {1} (\bibinfo {year} {1974})}\BibitemShut {NoStop}%
\bibitem [{\citenamefont {Fleury}\ and\ \citenamefont
  {Loudon}(1968)}]{fleury68}%
  \BibitemOpen
  \bibfield  {author} {\bibinfo {author} {\bibfnamefont {P.}~\bibnamefont
  {Fleury}}\ and\ \bibinfo {author} {\bibfnamefont {R.}~\bibnamefont
  {Loudon}},\ }\href {\doibase 10.1103/physrev.166.514} {\bibfield  {journal}
  {\bibinfo  {journal} {Phys. Rev.}\ }\textbf {\bibinfo {volume} {166}},\
  \bibinfo {pages} {514} (\bibinfo {year} {1968})}\BibitemShut {NoStop}%
\bibitem [{\citenamefont {Bouloux}\ \emph {et~al.}(1981)\citenamefont
  {Bouloux}, \citenamefont {Soubeyroux}, \citenamefont {Flem},\ and\
  \citenamefont {Hagenguller}}]{bouloux81}%
  \BibitemOpen
  \bibfield  {author} {\bibinfo {author} {\bibfnamefont {J.-C.}\ \bibnamefont
  {Bouloux}}, \bibinfo {author} {\bibfnamefont {J.-L.}\ \bibnamefont
  {Soubeyroux}}, \bibinfo {author} {\bibfnamefont {G.~L.}\ \bibnamefont
  {Flem}}, \ and\ \bibinfo {author} {\bibfnamefont {P.}~\bibnamefont
  {Hagenguller}},\ }\href {\doibase 10.1016/0022-4596(81)90469-2} {\bibfield
  {journal} {\bibinfo  {journal} {J. Solid State Chem.}\ }\textbf {\bibinfo
  {volume} {38}},\ \bibinfo {pages} {34} (\bibinfo {year} {1981})}\BibitemShut
  {NoStop}%
\bibitem [{Note2()}]{Note2}%
  \BibitemOpen
  \bibinfo {note} {A significant overestimation of the inter-site exchange
  within an approach based on Hubbard-I has been previously pointed out in
  Ref.~\protect \rev@citealpnum {Wan2006}. Apparently, neglecting hybridization
  effects in calculation of the impurity Green's function is a rather crude
  approximation even for strongly-localized Mott insulators.}\BibitemShut
  {Stop}%
\bibitem [{\citenamefont {Birgeneau}\ \emph {et~al.}(1970)\citenamefont
  {Birgeneau}, \citenamefont {Guggenheim},\ and\ \citenamefont
  {Shirane}}]{birgeneau70}%
  \BibitemOpen
  \bibfield  {author} {\bibinfo {author} {\bibfnamefont {R.}~\bibnamefont
  {Birgeneau}}, \bibinfo {author} {\bibfnamefont {H.}~\bibnamefont
  {Guggenheim}}, \ and\ \bibinfo {author} {\bibfnamefont {G.}~\bibnamefont
  {Shirane}},\ }\href
  {http://journals.aps.org/prb/abstract/10.1103/PhysRevB.1.2211} {\bibfield
  {journal} {\bibinfo  {journal} {Phys. Rev. B}\ }\textbf {\bibinfo {volume}
  {1}},\ \bibinfo {pages} {2211} (\bibinfo {year} {1970})}\BibitemShut
  {NoStop}%
\bibitem [{\citenamefont {Takeda}\ and\ \citenamefont
  {\={O}hara}(1974)}]{takeda74}%
  \BibitemOpen
  \bibfield  {author} {\bibinfo {author} {\bibfnamefont {T.}~\bibnamefont
  {Takeda}}\ and\ \bibinfo {author} {\bibfnamefont {S.}~\bibnamefont
  {\={O}hara}},\ }\href {\doibase 10.1143/JPSJ.37.275} {\bibfield  {journal}
  {\bibinfo  {journal} {J. Phys. Soc. Jpn.}\ }\textbf {\bibinfo {volume}
  {37}},\ \bibinfo {pages} {275} (\bibinfo {year} {1974})}\BibitemShut
  {NoStop}%
\bibitem [{\citenamefont {Kastner}\ \emph {et~al.}(1998)\citenamefont
  {Kastner}, \citenamefont {Birgeneau}, \citenamefont {Shirane},\ and\
  \citenamefont {Endoh}}]{kastner_rmp_1998}%
  \BibitemOpen
  \bibfield  {author} {\bibinfo {author} {\bibfnamefont {M.}~\bibnamefont
  {Kastner}}, \bibinfo {author} {\bibfnamefont {R.}~\bibnamefont {Birgeneau}},
  \bibinfo {author} {\bibfnamefont {G.}~\bibnamefont {Shirane}}, \ and\
  \bibinfo {author} {\bibfnamefont {Y.}~\bibnamefont {Endoh}},\ }\href
  {\doibase 10.1103/RevModPhys.70.897} {\bibfield  {journal} {\bibinfo
  {journal} {Rev. Mod. Phys.}\ }\textbf {\bibinfo {volume} {70}},\ \bibinfo
  {pages} {897} (\bibinfo {year} {1998})}\BibitemShut {NoStop}%
\bibitem [{\citenamefont {Dagotto}(1994)}]{dagotto_94}%
  \BibitemOpen
  \bibfield  {author} {\bibinfo {author} {\bibfnamefont {E.}~\bibnamefont
  {Dagotto}},\ }\href {\doibase 10.1103/RevModPhys.66.763} {\bibfield
  {journal} {\bibinfo  {journal} {Rev. Mod. Phys.}\ }\textbf {\bibinfo {volume}
  {66}},\ \bibinfo {pages} {763} (\bibinfo {year} {1994})}\BibitemShut
  {NoStop}%
\bibitem [{\citenamefont {Paglione}\ and\ \citenamefont
  {Greene}(2010)}]{paglione10}%
  \BibitemOpen
  \bibfield  {author} {\bibinfo {author} {\bibfnamefont {J.}~\bibnamefont
  {Paglione}}\ and\ \bibinfo {author} {\bibfnamefont {R.~L.}\ \bibnamefont
  {Greene}},\ }\href@noop {} {\bibfield  {journal} {\bibinfo  {journal} {Nature
  Phys.}\ }\textbf {\bibinfo {volume} {6}},\ \bibinfo {pages} {645} (\bibinfo
  {year} {2010})}\BibitemShut {NoStop}%
\bibitem [{\citenamefont {Hiley}\ \emph {et~al.}(2014)\citenamefont {Hiley},
  \citenamefont {Lees}, \citenamefont {Fisher}, \citenamefont {Thompsett},
  \citenamefont {Agrestini}, \citenamefont {Smith},\ and\ \citenamefont
  {Walton}}]{hiley14}%
  \BibitemOpen
  \bibfield  {author} {\bibinfo {author} {\bibfnamefont {C.~I.}\ \bibnamefont
  {Hiley}}, \bibinfo {author} {\bibfnamefont {M.~R.}\ \bibnamefont {Lees}},
  \bibinfo {author} {\bibfnamefont {J.~M.}\ \bibnamefont {Fisher}}, \bibinfo
  {author} {\bibfnamefont {D.}~\bibnamefont {Thompsett}}, \bibinfo {author}
  {\bibfnamefont {S.}~\bibnamefont {Agrestini}}, \bibinfo {author}
  {\bibfnamefont {R.~I.}\ \bibnamefont {Smith}}, \ and\ \bibinfo {author}
  {\bibfnamefont {R.~I.}\ \bibnamefont {Walton}},\ }\href {\doibase
  10.1002/ange.201310110} {\bibfield  {journal} {\bibinfo  {journal}
  {Angewandte Chemie}\ }\textbf {\bibinfo {volume} {126}},\ \bibinfo {pages}
  {4512} (\bibinfo {year} {2014})}\BibitemShut {NoStop}%
\bibitem [{\citenamefont {Seki}\ \emph {et~al.}(2014)\citenamefont {Seki},
  \citenamefont {Yamada}, \citenamefont {Saito}, \citenamefont {Kennedy},\ and\
  \citenamefont {Shimakawa}}]{seki14}%
  \BibitemOpen
  \bibfield  {author} {\bibinfo {author} {\bibfnamefont {H.}~\bibnamefont
  {Seki}}, \bibinfo {author} {\bibfnamefont {R.}~\bibnamefont {Yamada}},
  \bibinfo {author} {\bibfnamefont {T.}~\bibnamefont {Saito}}, \bibinfo
  {author} {\bibfnamefont {B.~J.}\ \bibnamefont {Kennedy}}, \ and\ \bibinfo
  {author} {\bibfnamefont {Y.}~\bibnamefont {Shimakawa}},\ }\href@noop {}
  {\bibfield  {journal} {\bibinfo  {journal} {Inorg. Chem.}\ }\textbf {\bibinfo
  {volume} {53}},\ \bibinfo {pages} {4579} (\bibinfo {year}
  {2014})}\BibitemShut {NoStop}%
\bibitem [{\citenamefont {Savrasov}\ and\ \citenamefont
  {Kotliar}(2004)}]{Savrasov2004}%
  \BibitemOpen
  \bibfield  {author} {\bibinfo {author} {\bibfnamefont {S.~Y.}\ \bibnamefont
  {Savrasov}}\ and\ \bibinfo {author} {\bibfnamefont {G.}~\bibnamefont
  {Kotliar}},\ }\href@noop {} {\bibfield  {journal} {\bibinfo  {journal} {Phys.
  Rev. B}\ }\textbf {\bibinfo {volume} {69}},\ \bibinfo {pages} {245101}
  (\bibinfo {year} {2004})}\BibitemShut {NoStop}%
\bibitem [{\citenamefont {Pourovskii}()}]{pourovskii14u}%
  \BibitemOpen
  \bibfield  {author} {\bibinfo {author} {\bibfnamefont {L.}~\bibnamefont
  {Pourovskii}},\ }\href@noop {} {}\bibinfo {note} {In preparation}\BibitemShut
  {NoStop}%
\end{thebibliography}

\end{document}